%
%
%
%
%
%
%
%
\def\standardrisposta{s }\def\reducedrisposta{r }
\def\mplarisposta{mpla }\def\zerorisposta{z }
\def\doublerisposta{d }\def\cartarisposta{e }\def\amsrisposta{y }
\newcount\ingrandimento \newcount\sinnota \newcount\dimnota
\newcount\unoduecol \newdimen\collhsize \newdimen\tothsize
\newdimen\fullhsize \newcount\controllorisposta \sinnota=1
\newskip\infralinea  \global\controllorisposta=0
%
%
%
%
%
\def\risposta{s }
\def\srisposta{e }
\def\arisposta{y }
\ifx\risposta\standardrisposta \ingrandimento=1200
\message {>> This will come out UNREDUCED << }
\dimnota=2 \unoduecol=1 \global\controllorisposta=1 \fi
\ifx\risposta\reducedrisposta \ingrandimento=1095 \dimnota=1
\unoduecol=1  \global\controllorisposta=1
\message {>> This will come out REDUCED << } \fi
\ifx\risposta\doublerisposta \ingrandimento=1000 \dimnota=2
\unoduecol=2   \message {>> You must print this in
LANDSCAPE orientation << } \global\controllorisposta=1 \fi
\ifx\risposta\mplarisposta \ingrandimento=1000 \dimnota=1
\message {>> Mod. Phys. Lett. A format << }
\unoduecol=1 \global\controllorisposta=1 \fi
\ifx\risposta\zerorisposta \ingrandimento=1000 \dimnota=2
\message {>> Zero Magnification format << }
\unoduecol=1 \global\controllorisposta=1 \fi
\ifnum\controllorisposta=0  \ingrandimento=1200
\message {>>> ERROR IN INPUT, I ASSUME STANDARD
UNREDUCED FORMAT <<< }  \dimnota=2 \unoduecol=1 \fi
\magnification=\ingrandimento
%
%
%
%
\newdimen\eucolumnsize \newdimen\eudoublehsize \newdimen\eudoublevsize
\newdimen\uscolumnsize \newdimen\usdoublehsize \newdimen\usdoublevsize
\newdimen\eusinglehsize \newdimen\eusinglevsize \newdimen\ussinglehsize
\newskip\standardbaselineskip \newdimen\ussinglevsize
\newskip\reducedbaselineskip \newskip\doublebaselineskip
\eucolumnsize=12.0truecm    
\eudoublehsize=25.5truecm   
\eudoublevsize=6.5truein    
\uscolumnsize=4.4truein     
\usdoublehsize=9.4truein    
\usdoublevsize=6.8truein    
\eusinglehsize=6.5truein    
\eusinglevsize=24truecm     
\ussinglehsize=6.5truein    
\ussinglevsize=8.9truein    
\standardbaselineskip=16pt plus.2pt  
\reducedbaselineskip=14pt plus.2pt   
\doublebaselineskip=12pt plus.2pt    
%
%
\def\Portoffset{}
\def\Landoffset{}
\ifx\risposta\mplarisposta \def\Portoffset{\hoffset=1.8truecm} \fi
%
%
\def\Landspec{}
\tolerance=10000
\parskip=0pt plus2pt  \leftskip=0pt \rightskip=0pt
%
%
\ifx\risposta\standardrisposta \infralinea=\standardbaselineskip \fi
\ifx\risposta\reducedrisposta  \infralinea=\reducedbaselineskip \fi
\ifx\risposta\doublerisposta   \infralinea=\doublebaselineskip \fi
\ifx\risposta\mplarisposta     \infralinea=13pt \fi
\ifx\risposta\zerorisposta     \infralinea=12pt plus.2pt\fi
\ifnum\controllorisposta=0    \infralinea=\standardbaselineskip \fi
\ifx\risposta\doublerisposta   \Landoffset \else \Portoffset \fi
\ifx\risposta\doublerisposta \ifx\srisposta\cartarisposta
\tothsize=\eudoublehsize \collhsize=\eucolumnsize
\vsize=\eudoublevsize  \else  \tothsize=\usdoublehsize
\collhsize=\uscolumnsize \vsize=\usdoublevsize \fi \else
\ifx\srisposta\cartarisposta \tothsize=\eusinglehsize
\vsize=\eusinglevsize \else  \tothsize=\ussinglehsize
\vsize=\ussinglevsize \fi \collhsize=4.4truein \fi
\ifx\risposta\mplarisposta \tothsize=5.0truein
\vsize=7.8truein \collhsize=4.4truein \fi
%
%
%
%
\newcount\contaeuler \newcount\contacyrill \newcount\contaams
\font\ninerm=cmr9  \font\eightrm=cmr8  \font\sixrm=cmr6
\font\ninei=cmmi9  \font\eighti=cmmi8  \font\sixi=cmmi6
\font\ninesy=cmsy9  \font\eightsy=cmsy8  \font\sixsy=cmsy6
\font\ninebf=cmbx9  \font\eightbf=cmbx8  \font\sixbf=cmbx6
\font\ninett=cmtt9  \font\eighttt=cmtt8  \font\nineit=cmti9
\font\eightit=cmti8 \font\ninesl=cmsl9  \font\eightsl=cmsl8
\skewchar\ninei='177 \skewchar\eighti='177 \skewchar\sixi='177
\skewchar\ninesy='60 \skewchar\eightsy='60 \skewchar\sixsy='60
\hyphenchar\ninett=-1 \hyphenchar\eighttt=-1 \hyphenchar\tentt=-1
\def\bfmath{\cmmib}                 
\font\tencmmib=cmmib10  \newfam\cmmibfam  \skewchar\tencmmib='177
\font\tencmbsy=cmbsy10  \newfam\cmbsyfam  \skewchar\tencmbsy='60
\def\scaps{\cmcsc}                 
\font\tencmcsc=cmcsc10  \newfam\cmcscfam
\ifnum\ingrandimento=1095

\font\capsone=cmcsc10 at 10.95pt 

\else

\font\capsone=cmcsc10 at 12pt 
\fi

\def\ttaarr{\bf}		
\def\ppaarr{\sl}		

%
%
%
\newfam\eufmfam \newfam\msamfam \newfam\msbmfam \newfam\eufbfam
\def\Loadeulerfonts{\global\contaeuler=1 \ifx\arisposta\amsrisposta
\font\teneufm=eufm10              
\font\eighteufm=eufm8 \font\nineeufm=eufm9 \font\sixeufm=eufm6
\font\seveneufm=eufm7  \font\fiveeufm=eufm5
\font\teneufb=eufb10              
\font\eighteufb=eufb8 \font\nineeufb=eufb9 \font\sixeufb=eufb6
\font\seveneufb=eufb7  \font\fiveeufb=eufb5
\font\teneurm=eurm10              
\font\eighteurm=eurm8 \font\nineeurm=eurm9
\font\teneurb=eurb10              
\font\eighteurb=eurb8 \font\nineeurb=eurb9
\font\teneusm=eusm10              
\font\eighteusm=eusm8 \font\nineeusm=eusm9
\font\teneusb=eusb10              
\font\eighteusb=eusb8 \font\nineeusb=eusb9
\else \def\eufm{\tt} \def\eufb{\tt} \def\eurm{\tt} \def\eurb{\tt}
\def\eusm{\tt} \def\eusb{\tt}    \fi}
\def\loadeuler{\Loadeulerfonts\tenpoint}
\def\loadamsmath{\global\contaams=1 \ifx\arisposta\amsrisposta
\font\tenmsam=msam10 \font\ninemsam=msam9 \font\eightmsam=msam8
\font\sevenmsam=msam7 \font\sixmsam=msam6 \font\fivemsam=msam5
\font\tenmsbm=msbm10 \font\ninemsbm=msbm9 \font\eightmsbm=msbm8
\font\sevenmsbm=msbm7 \font\sixmsbm=msbm6 \font\fivemsbm=msbm5
\else \def\msbm{\bf} \fi \def\Bbb{\msbm} \def\symbl{\msam} \tenpoint}
\def\loadcyrill{\global\contacyrill=1 \ifx\arisposta\amsrisposta
\font\tenwncyr=wncyr10 \font\ninewncyr=wncyr9 \font\eightwncyr=wncyr8
\font\tenwncyb=wncyr10 \font\ninewncyb=wncyr9 \font\eightwncyb=wncyr8
\font\tenwncyi=wncyr10 \font\ninewncyi=wncyr9 \font\eightwncyi=wncyr8
\else \def\cyrill{\sl} \def\cyrilb{\sl} \def\cyrili{\sl} \fi\tenpoint}
\ifx\arisposta\amsrisposta
\font\sevenex=cmex7               
\font\eightex=cmex8  \font\nineex=cmex9
\font\ninecmmib=cmmib9   \font\eightcmmib=cmmib8
\font\sevencmmib=cmmib7 \font\sixcmmib=cmmib6
\font\fivecmmib=cmmib5   \skewchar\ninecmmib='177
\skewchar\eightcmmib='177  \skewchar\sevencmmib='177
\skewchar\sixcmmib='177   \skewchar\fivecmmib='177
\font\ninecmbsy=cmbsy9    \font\eightcmbsy=cmbsy8
\font\sevencmbsy=cmbsy7  \font\sixcmbsy=cmbsy6
\font\fivecmbsy=cmbsy5   \skewchar\ninecmbsy='60
\skewchar\eightcmbsy='60  \skewchar\sevencmbsy='60
\skewchar\sixcmbsy='60    \skewchar\fivecmbsy='60
\font\ninecmcsc=cmcsc9    \font\eightcmcsc=cmcsc8     \else
\def\cmmib{\fam\cmmibfam\tencmmib}\textfont\cmmibfam=\tencmmib
\scriptfont\cmmibfam=\tencmmib \scriptscriptfont\cmmibfam=\tencmmib
\def\cmbsy{\fam\cmbsyfam\tencmbsy} \textfont\cmbsyfam=\tencmbsy
\scriptfont\cmbsyfam=\tencmbsy \scriptscriptfont\cmbsyfam=\tencmbsy
\scriptfont\cmcscfam=\tencmcsc \scriptscriptfont\cmcscfam=\tencmcsc
\def\cmcsc{\fam\cmcscfam\tencmcsc} \textfont\cmcscfam=\tencmcsc \fi
\catcode`@=11
\newskip\ttglue
\gdef\tenpoint{\def\rm{\fam0\tenrm}
  \textfont0=\tenrm \scriptfont0=\sevenrm \scriptscriptfont0=\fiverm
  \textfont1=\teni \scriptfont1=\seveni \scriptscriptfont1=\fivei
  \textfont2=\tensy \scriptfont2=\sevensy \scriptscriptfont2=\fivesy
  \textfont3=\tenex \scriptfont3=\tenex \scriptscriptfont3=\tenex
  \def\mcal{\fam2 \tensy}  \def\mmit{\fam1 \teni}
  \textfont\itfam=\tenit \def\it{\fam\itfam\tenit}
  \textfont\slfam=\tensl \def\sl{\fam\slfam\tensl}
  \textfont\ttfam=\tentt \scriptfont\ttfam=\eighttt
  \scriptscriptfont\ttfam=\eighttt  \def\tt{\fam\ttfam\tentt}
  \textfont\bffam=\tenbf \scriptfont\bffam=\sevenbf
  \scriptscriptfont\bffam=\fivebf \def\bf{\fam\bffam\tenbf}
     \ifx\arisposta\amsrisposta    \ifnum\contaeuler=1
  \textfont\eufmfam=\teneufm \scriptfont\eufmfam=\seveneufm
  \scriptscriptfont\eufmfam=\fiveeufm \def\eufm{\fam\eufmfam\teneufm}
  \textfont\eufbfam=\teneufb \scriptfont\eufbfam=\seveneufb
  \scriptscriptfont\eufbfam=\fiveeufb \def\eufb{\fam\eufbfam\teneufb}
  \def\eurm{\teneurm} \def\eurb{\teneurb} \def\eusm{\teneusm}
  \def\eusb{\teneusb}    \fi    \ifnum\contaams=1
  \textfont\msamfam=\tenmsam \scriptfont\msamfam=\sevenmsam
  \scriptscriptfont\msamfam=\fivemsam \def\msam{\fam\msamfam\tenmsam}
  \textfont\msbmfam=\tenmsbm \scriptfont\msbmfam=\sevenmsbm
  \scriptscriptfont\msbmfam=\fivemsbm \def\msbm{\fam\msbmfam\tenmsbm}
     \fi      \ifnum\contacyrill=1     \def\cyrill{\tenwncyr}
  \def\cyrilb{\tenwncyb}  \def\cyrili{\tenwncyi}         \fi
  \textfont3=\tenex \scriptfont3=\sevenex \scriptscriptfont3=\sevenex
  \def\cmmib{\fam\cmmibfam\tencmmib} \scriptfont\cmmibfam=\sevencmmib
  \textfont\cmmibfam=\tencmmib  \scriptscriptfont\cmmibfam=\fivecmmib
  \def\cmbsy{\fam\cmbsyfam\tencmbsy} \scriptfont\cmbsyfam=\sevencmbsy
  \textfont\cmbsyfam=\tencmbsy  \scriptscriptfont\cmbsyfam=\fivecmbsy
  \def\cmcsc{\fam\cmcscfam\tencmcsc} \scriptfont\cmcscfam=\eightcmcsc
  \textfont\cmcscfam=\tencmcsc \scriptscriptfont\cmcscfam=\eightcmcsc
     \fi            \tt \ttglue=.5em plus.25em minus.15em
  \normalbaselineskip=12pt
  \setbox\strutbox=\hbox{\vrule height8.5pt depth3.5pt width0pt}
  \let\sc=\eightrm \let\big=\tenbig   \normalbaselines
  \baselineskip=\infralinea  \rm}
\gdef\ninepoint{\def\rm{\fam0\ninerm}
  \textfont0=\ninerm \scriptfont0=\sixrm \scriptscriptfont0=\fiverm
  \textfont1=\ninei \scriptfont1=\sixi \scriptscriptfont1=\fivei
  \textfont2=\ninesy \scriptfont2=\sixsy \scriptscriptfont2=\fivesy
  \textfont3=\tenex \scriptfont3=\tenex \scriptscriptfont3=\tenex
  \def\mcal{\fam2 \ninesy}  \def\mmit{\fam1 \ninei}
  \textfont\itfam=\nineit \def\it{\fam\itfam\nineit}
  \textfont\slfam=\ninesl \def\sl{\fam\slfam\ninesl}
  \textfont\ttfam=\ninett \scriptfont\ttfam=\eighttt
  \scriptscriptfont\ttfam=\eighttt \def\tt{\fam\ttfam\ninett}
  \textfont\bffam=\ninebf \scriptfont\bffam=\sixbf
  \scriptscriptfont\bffam=\fivebf \def\bf{\fam\bffam\ninebf}
     \ifx\arisposta\amsrisposta  \ifnum\contaeuler=1
  \textfont\eufmfam=\nineeufm \scriptfont\eufmfam=\sixeufm
  \scriptscriptfont\eufmfam=\fiveeufm \def\eufm{\fam\eufmfam\nineeufm}
  \textfont\eufbfam=\nineeufb \scriptfont\eufbfam=\sixeufb
  \scriptscriptfont\eufbfam=\fiveeufb \def\eufb{\fam\eufbfam\nineeufb}
  \def\eurm{\nineeurm} \def\eurb{\nineeurb} \def\eusm{\nineeusm}
  \def\eusb{\nineeusb}     \fi   \ifnum\contaams=1
  \textfont\msamfam=\ninemsam \scriptfont\msamfam=\sixmsam
  \scriptscriptfont\msamfam=\fivemsam \def\msam{\fam\msamfam\ninemsam}
  \textfont\msbmfam=\ninemsbm \scriptfont\msbmfam=\sixmsbm
  \scriptscriptfont\msbmfam=\fivemsbm \def\msbm{\fam\msbmfam\ninemsbm}
     \fi       \ifnum\contacyrill=1     \def\cyrill{\ninewncyr}
  \def\cyrilb{\ninewncyb}  \def\cyrili{\ninewncyi}         \fi
  \textfont3=\nineex \scriptfont3=\sevenex \scriptscriptfont3=\sevenex
  \def\cmmib{\fam\cmmibfam\ninecmmib}  \textfont\cmmibfam=\ninecmmib
  \scriptfont\cmmibfam=\sixcmmib \scriptscriptfont\cmmibfam=\fivecmmib
  \def\cmbsy{\fam\cmbsyfam\ninecmbsy}  \textfont\cmbsyfam=\ninecmbsy
  \scriptfont\cmbsyfam=\sixcmbsy \scriptscriptfont\cmbsyfam=\fivecmbsy
  \def\cmcsc{\fam\cmcscfam\ninecmcsc} \scriptfont\cmcscfam=\eightcmcsc
  \textfont\cmcscfam=\ninecmcsc \scriptscriptfont\cmcscfam=\eightcmcsc
     \fi            \tt \ttglue=.5em plus.25em minus.15em
  \normalbaselineskip=11pt
  \setbox\strutbox=\hbox{\vrule height8pt depth3pt width0pt}
  \let\sc=\sevenrm \let\big=\ninebig \normalbaselines\rm}
\gdef\eightpoint{\def\rm{\fam0\eightrm}
  \textfont0=\eightrm \scriptfont0=\sixrm \scriptscriptfont0=\fiverm
  \textfont1=\eighti \scriptfont1=\sixi \scriptscriptfont1=\fivei
  \textfont2=\eightsy \scriptfont2=\sixsy \scriptscriptfont2=\fivesy
  \textfont3=\tenex \scriptfont3=\tenex \scriptscriptfont3=\tenex
  \def\mcal{\fam2 \eightsy}  \def\mmit{\fam1 \eighti}
  \textfont\itfam=\eightit \def\it{\fam\itfam\eightit}
  \textfont\slfam=\eightsl \def\sl{\fam\slfam\eightsl}
  \textfont\ttfam=\eighttt \scriptfont\ttfam=\eighttt
  \scriptscriptfont\ttfam=\eighttt \def\tt{\fam\ttfam\eighttt}
  \textfont\bffam=\eightbf \scriptfont\bffam=\sixbf
  \scriptscriptfont\bffam=\fivebf \def\bf{\fam\bffam\eightbf}
     \ifx\arisposta\amsrisposta   \ifnum\contaeuler=1
  \textfont\eufmfam=\eighteufm \scriptfont\eufmfam=\sixeufm
  \scriptscriptfont\eufmfam=\fiveeufm \def\eufm{\fam\eufmfam\eighteufm}
  \textfont\eufbfam=\eighteufb \scriptfont\eufbfam=\sixeufb
  \scriptscriptfont\eufbfam=\fiveeufb \def\eufb{\fam\eufbfam\eighteufb}
  \def\eurm{\eighteurm} \def\eurb{\eighteurb} \def\eusm{\eighteusm}
  \def\eusb{\eighteusb}       \fi    \ifnum\contaams=1
  \textfont\msamfam=\eightmsam \scriptfont\msamfam=\sixmsam
  \scriptscriptfont\msamfam=\fivemsam \def\msam{\fam\msamfam\eightmsam}
  \textfont\msbmfam=\eightmsbm \scriptfont\msbmfam=\sixmsbm
  \scriptscriptfont\msbmfam=\fivemsbm \def\msbm{\fam\msbmfam\eightmsbm}
     \fi       \ifnum\contacyrill=1     \def\cyrill{\eightwncyr}
  \def\cyrilb{\eightwncyb}  \def\cyrili{\eightwncyi}         \fi
  \textfont3=\eightex \scriptfont3=\sevenex \scriptscriptfont3=\sevenex
  \def\cmmib{\fam\cmmibfam\eightcmmib}  \textfont\cmmibfam=\eightcmmib
  \scriptfont\cmmibfam=\sixcmmib \scriptscriptfont\cmmibfam=\fivecmmib
  \def\cmbsy{\fam\cmbsyfam\eightcmbsy}  \textfont\cmbsyfam=\eightcmbsy
  \scriptfont\cmbsyfam=\sixcmbsy \scriptscriptfont\cmbsyfam=\fivecmbsy
  \def\cmcsc{\fam\cmcscfam\eightcmcsc} \scriptfont\cmcscfam=\eightcmcsc
  \textfont\cmcscfam=\eightcmcsc \scriptscriptfont\cmcscfam=\eightcmcsc
     \fi             \tt \ttglue=.5em plus.25em minus.15em
  \normalbaselineskip=9pt
  \setbox\strutbox=\hbox{\vrule height7pt depth2pt width0pt}
  \let\sc=\sixrm \let\big=\eightbig \normalbaselines\rm }
\gdef\tenbig#1{{\hbox{$\left#1\vbox to8.5pt{}\right.\n@space$}}}
\gdef\ninebig#1{{\hbox{$\textfont0=\tenrm\textfont2=\tensy
   \left#1\vbox to7.25pt{}\right.\n@space$}}}
\gdef\eightbig#1{{\hbox{$\textfont0=\ninerm\textfont2=\ninesy
   \left#1\vbox to6.5pt{}\right.\n@space$}}}
\def\alternativefont#1#2{\ifx\arisposta\amsrisposta \relax \else
\xdef#1{#2} \fi}
\global\contaeuler=0 \global\contacyrill=0 \global\contaams=0
%
%
%
%
\newbox\fotlinebb \newbox\hedlinebb \newbox\leftcolumn
\gdef\makeheadline{\vbox to 0pt{\vskip-22.5pt
     \fullline{\vbox to8.5pt{}\the\headline}\vss}\nointerlineskip}
\gdef\makehedlinebb{\vbox to 0pt{\vskip-22.5pt
     \fullline{\vbox to8.5pt{}\copy\hedlinebb\hfil
     \line{\hfill\the\headline\hfill}}\vss} \nointerlineskip}
\gdef\makefootline{\baselineskip=24pt \fullline{\the\footline}}
\gdef\makefotlinebb{\baselineskip=24pt
    \fullline{\copy\fotlinebb\hfil\line{\hfill\the\footline\hfill}}}
\gdef\doubleformat{\shipout\vbox{\Landspec\makehedlinebb
     \fullline{\box\leftcolumn\hfil\columnbox}\makefotlinebb}
     \advancepageno}
\gdef\columnbox{\leftline{\pagebody}}
\gdef\line#1{\hbox to\hsize{\hskip\leftskip#1\hskip\rightskip}}
\gdef\fullline#1{\hbox to\fullhsize{\hskip\leftskip{#1}%
\hskip\rightskip}}
\gdef\footnote#1{\let\@sf=\empty
         \ifhmode\edef\#sf{\spacefactor=\the\spacefactor}\/\fi
         #1\@sf\vfootnote{#1}}
\gdef\vfootnote#1{\insert\footins\bgroup
         \ifnum\dimnota=1  \eightpoint\fi
         \ifnum\dimnota=2  \ninepoint\fi
         \ifnum\dimnota=0  \tenpoint\fi
         \interlinepenalty=\interfootnotelinepenalty
         \splittopskip=\ht\strutbox
         \splitmaxdepth=\dp\strutbox \floatingpenalty=20000
         \leftskip=\oldssposta \rightskip=\olddsposta
         \spaceskip=0pt \xspaceskip=0pt
         \ifnum\sinnota=0   \textindent{#1}\fi
         \ifnum\sinnota=1   \item{#1}\fi
         \footstrut\futurelet\next\fo@t}
\gdef\fo@t{\ifcat\bgroup\noexpand\next \let\next\f@@t
             \else\let\next\f@t\fi \next}
\gdef\f@@t{\bgroup\aftergroup\@foot\let\next}
\gdef\f@t#1{#1\@foot} \gdef\@foot{\strut\egroup}
\gdef\footstrut{\vbox to\splittopskip{}}
\skip\footins=\bigskipamount
\count\footins=1000  \dimen\footins=8in
\catcode`@=12
\tenpoint
\ifnum\unoduecol=1 \hsize=\tothsize   \fullhsize=\tothsize \fi
\ifnum\unoduecol=2 \hsize=\collhsize  \fullhsize=\tothsize \fi
\global\let\lrcol=L      \ifnum\unoduecol=1
\output{\plainoutput{\ifnum\tipbnota=2 \clearnmbnota\fi}} \fi
\ifnum\unoduecol=2 \output{\if L\lrcol
     \global\setbox\leftcolumn=\columnbox
     \global\setbox\fotlinebb=\line{\hfill\the\footline\hfill}
     \global\setbox\hedlinebb=\line{\hfill\the\headline\hfill}
     \advancepageno  \global\let\lrcol=R
     \else  \doubleformat \global\let\lrcol=L \fi
     \ifnum\outputpenalty>-20000 \else\dosupereject\fi
     \ifnum\tipbnota=2\clearnmbnota\fi }\fi
\def\ifdoublepage{\ifnum\unoduecol=2 }
\gdef\yespagenumbers{\footline={\hss\tenrm\folio\hss}}
\gdef\ciao{ \ifnum\fdefcontre=1 \endfdef\fi
     \par\vfill\supereject \ifnum\unoduecol=2
     \if R\lrcol  \headline={}\nopagenumbers\null\vfill\eject
     \fi\fi \end}

\newskip\olddsposta \newskip\oldssposta
\global\oldssposta=\leftskip \global\olddsposta=\rightskip

\def\filldots{\leaders\hbox to 1em{\hss.\hss}\hfill}
\def\inquadrb#1 {\vbox {\hrule  \hbox{\vrule \vbox {\vskip .2cm
    \hbox {\ #1\ } \vskip .2cm } \vrule  }  \hrule} }
 \def\newline{\hfil\break}
\def\jump{\vskip\baselineskip} \newskip\iinnffrr
\def\sjump{\iinnffrr=\baselineskip
          \divide\iinnffrr by 2 \vskip\iinnffrr}
\def\bjump{\vskip\baselineskip \vskip\baselineskip}
\newcount\nmbnota  \def\clearnmbnota{\global\nmbnota=0}
\newcount\tipbnota \def\letterfootnote{\global\tipbnota=1}

\def\note#1{\global\advance\nmbnota by 1 \ifnum\tipbnota=1
    \footnote{$^{\rm\nttlett}$}{#1} \else {\ifnum\tipbnota=2
    \footnote{$^{\nttsymb}$}{#1}
    \else\footnote{$^{\the\nmbnota}$}{#1}\fi}\fi}
\def\nttlett{\ifcase\nmbnota \or a\or b\or c\or d\or e\or f\or
g\or h\or i\or j\or k\or l\or m\or n\or o\or p\or q\or r\or
s\or t\or u\or v\or w\or y\or x\or z\fi}
\def\nttsymb{\ifcase\nmbnota \or\dag\or\sharp\or\ddag\or\star\or
\natural\or\flat\or\clubsuit\or\diamondsuit\or\heartsuit
\or\spadesuit\fi}   \clearnmbnota
\def\numberfootnote{\global\tipbnota=0} \numberfootnote
\def\setnote#1{\expandafter\xdef\csname#1\endcsname{
\ifnum\tipbnota=1 {\rm\nttlett} \else {\ifnum\tipbnota=2
{\nttsymb} \else \the\nmbnota\fi}\fi} }
\newcount\nbmfig  \def\clearnbmfig{\global\nbmfig=0}
\gdef\figure{\global\advance\nbmfig by 1
      {\rm fig. \the\nbmfig}}   \clearnbmfig
\def\setfig#1{\expandafter\xdef\csname#1\endcsname{fig. \the\nbmfig}}
 \def\endformula{\eqno\numero $$}
 \def\efr{\endformula}
\newcount\frmcount \def\clearfrmcount{\global\frmcount=0}
\def\numero{\global\advance\frmcount by 1   \ifnum\indappcount=0
  {\ifnum\cpcount <1 {\hbox{\rm (\the\frmcount )}}  \else
  {\hbox{\rm (\the\cpcount .\the\frmcount )}} \fi}  \else
  {\hbox{\rm (\applett .\the\frmcount )}} \fi}
\def\nameformula#1{\global\advance\frmcount by 1%
\ifnum\draftnum=0  {\ifnum\indappcount=0%
{\ifnum\cpcount<1\xdef\spzzttrra{(\the\frmcount )}%
\else\xdef\spzzttrra{(\the\cpcount .\the\frmcount )}\fi}%
\else\xdef\spzzttrra{(\applett .\the\frmcount )}\fi}%
\else\xdef\spzzttrra{(#1)}\fi%
\expandafter\xdef\csname#1\endcsname{\spzzttrra}
\eqno \hbox{\rm\spzzttrra} $$}
\def\nfr{\nameformula}    
\def\nameali#1{\global\advance\frmcount by 1%
\ifnum\draftnum=0  {\ifnum\indappcount=0%
{\ifnum\cpcount<1\xdef\spzzttrra{(\the\frmcount )}%
\else\xdef\spzzttrra{(\the\cpcount .\the\frmcount )}\fi}%
\else\xdef\spzzttrra{(\applett .\the\frmcount )}\fi}%
\else\xdef\spzzttrra{(#1)}\fi%
\expandafter\xdef\csname#1\endcsname{\spzzttrra}
  \hbox{\rm\spzzttrra} }      \clearfrmcount
\newcount\cpcount \def\clearcpcount{\global\cpcount=0}
\newcount\subcpcount \def\clearsubcpcount{\global\subcpcount=0}
\newcount\appcount \def\clearappcount{\global\appcount=0}
\newcount\indappcount \def\clearindappcount{\indappcount=0}
\newcount\sottoparcount 

\def\applett{\ifcase\appcount  \or {A}\or {B}\or {C}\or
{D}\or {E}\or {F}\or {G}\or {H}\or {I}\or {J}\or {K}\or {L}\or
{M}\or {N}\or {O}\or {P}\or {Q}\or {R}\or {S}\or {T}\or {U}\or
{V}\or {W}\or {X}\or {Y}\or {Z}\fi    \ifnum\appcount<0
\immediate\write16 {Panda ERROR - Appendix: counter "appcount"
out of range}\fi  \ifnum\appcount>26  \immediate\write16 {Panda
ERROR - Appendix: counter "appcount" out of range}\fi}
\clearappcount  \clearindappcount \newcount\connttrre
\def\clearconnttrre{\global\connttrre=0} \newcount\countref
\def\clearcountref{\global\countref=0} \clearcountref
\def\chapter#1{\global\advance\cpcount by 1 \clearfrmcount
                 \goodbreak\null\vbox{\jump\nobreak
                 \clearsubcpcount\clearindappcount
                 \itemitem{\ttaarr\the\cpcount .\qquad}{\ttaarr #1}
                 \par\nobreak\jump\sjump}\nobreak}
\def\section#1{\global\advance\subcpcount by 1 \goodbreak\null
               \vbox{\sjump\nobreak\ifnum\indappcount=0
                 {\ifnum\cpcount=0 {\itemitem{\ppaarr
               .\the\subcpcount\quad\enskip\ }{\ppaarr #1}\par} \else
                 {\itemitem{\ppaarr\the\cpcount .\the\subcpcount\quad
                  \enskip\ }{\ppaarr #1} \par}  \fi}
                \else{\itemitem{\ppaarr\applett .\the\subcpcount\quad
                 \enskip\ }{\ppaarr #1}\par}\fi\nobreak\jump}\nobreak}
\clearsubcpcount
\def\appendix#1{\global\advance\appcount by 1 \clearfrmcount
                  \goodbreak\null\vbox{\jump\nobreak
                  \global\advance\indappcount by 1 \clearsubcpcount
          \itemitem{ }{\hskip-40pt\ttaarr Appendix\ #1}
             \nobreak\jump\sjump}\nobreak}
\clearappcount \clearindappcount
\def\references{\goodbreak\null\vbox{\jump\nobreak
   \itemitem{}{\ttaarr References} \nobreak\jump\sjump}\nobreak}

\clearcpcount\clearcountref

\def\setchap#1{\ifnum\indappcount=0{\ifnum\subcpcount=0%
\xdef\spzzttrra{\the\cpcount}%
\else\xdef\spzzttrra{\the\cpcount .\the\subcpcount}\fi}
\else{\ifnum\subcpcount=0 \xdef\spzzttrra{\applett}%
\else\xdef\spzzttrra{\applett .\the\subcpcount}\fi}\fi
\expandafter\xdef\csname#1\endcsname{\spzzttrra}}
\newcount\draftnum \newcount\ppora   \newcount\ppminuti
\global\ppora=\time   \global\ppminuti=\time
\global\divide\ppora by 60  \draftnum=\ppora
\multiply\draftnum by 60    \global\advance\ppminuti by -\draftnum
\def\droggi{\number\day /\number\month /\number\year\ \the\ppora
:\the\ppminuti}     \global\draftnum=0
\def\draftcomment#1{\ifnum\draftnum=0 \relax \else
{\ {\bf ***}\ #1\ {\bf ***}\ }\fi} 
%
%
\catcode`@=11
\gdef\Ref#1{\expandafter\ifx\csname @rrxx@#1\endcsname\relax%
{\global\advance\countref by 1    \ifnum\countref>200
\immediate\write16 {Panda ERROR - Ref: maximum number of references
exceeded}  \expandafter\xdef\csname @rrxx@#1\endcsname{0}\else
\expandafter\xdef\csname @rrxx@#1\endcsname{\the\countref}\fi}\fi
\ifnum\draftnum=0 \csname @rrxx@#1\endcsname \else#1\fi}
\gdef\beginref{\ifnum\draftnum=0  \gdef\Rref{\fairef}
\gdef\endref{\scriviref} \else\relax\fi
\ifx\risposta\mplarisposta \ninepoint \fi
\parskip 2pt plus.2pt \baselineskip=12pt}
\def\Reflab#1{[#1]} \gdef\Rref#1#2{\item{\Reflab{#1}}{#2}}
\gdef\endref{\relax}  \newcount\conttemp
\gdef\fairef#1#2{\expandafter\ifx\csname @rrxx@#1\endcsname\relax
{\global\conttemp=0 \immediate\write16 {Panda ERROR - Ref: reference
[#1] undefined}} \else
{\global\conttemp=\csname @rrxx@#1\endcsname } \fi
\global\advance\conttemp by 50  \global\setbox\conttemp=\hbox{#2} }
\gdef\scriviref{\clearconnttrre\conttemp=50
\loop\ifnum\connttrre<\countref \advance\conttemp by 1
\advance\connttrre by 1
\item{\Reflab{\the\connttrre}}{\unhcopy\conttemp} \repeat}
\clearcountref \clearconnttrre
\catcode`@=12
\ifx\risposta\mplarisposta \def\Reflab#1{#1.} \letterfootnote \fi

\def\slashchar#1{\setbox0=\hbox{$#1$} \dimen0=\wd0
     \setbox1=\hbox{/} \dimen1=\wd1 \ifdim\dimen0>\dimen1
      \rlap{\hbox to \dimen0{\hfil/\hfil}} #1 \else
      \rlap{\hbox to \dimen1{\hfil$#1$\hfil}} / \fi}
\ifx\oldchi\undefined \let\oldchi=\chi
  \def\cchi{{\raise 1pt\hbox{$\oldchi$}}} \let\chi=\cchi \fi

\def\frac#1#2{{\textstyle{#1 \over #2}}}

\def\half{\ifinner {\scriptstyle {1 \over 2}}\else {1 \over 2} \fi}

\def\simge{\rlap{\raise 2pt \hbox{$>$}}{\lower 2pt \hbox{$\sim$}}}
\def\simle{\rlap{\raise 2pt \hbox{$<$}}{\lower 2pt \hbox{$\sim$}}}

\def\vbig#1#2{{\vbigd@men=#2\divide\vbigd@men by 2%
\hbox{$\left#1\vbox to \vbigd@men{}\right.\n@space$}}}

%
%
\newcount\fdefcontre \newcount\fdefcount \newcount\indcount
\newread\filefdef  \newread\fileftmp  \newwrite\filefdef
\newwrite\fileftmp     \def\strip#1*.A {#1}
\def\futuredef#1{\beginfdef
\expandafter\ifx\csname#1\endcsname\relax%
{\immediate\write\fileftmp {#1*.A}
\immediate\write16 {Panda Warning - fdef: macro "#1" on page
\the\pageno \space undefined}
\ifnum\draftnum=0 \expandafter\xdef\csname#1\endcsname{(?)}
\else \expandafter\xdef\csname#1\endcsname{(#1)} \fi
\global\advance\fdefcount by 1}\fi   \csname#1\endcsname}

\def\beginfdef{\ifnum\fdefcontre=0
\immediate\openin\filefdef \jobname.fdef
\immediate\openout\fileftmp \jobname.ftmp
\global\fdefcontre=1  \ifeof\filefdef \immediate\write16 {Panda
WARNING - fdef: file \jobname.fdef not found, run TeX again}
\else \immediate\read\filefdef to\spzzttrra
\global\advance\fdefcount by \spzzttrra
\indcount=0      \loop\ifnum\indcount<\fdefcount
\advance\indcount by 1   \immediate\read\filefdef to\spezttrra
\immediate\read\filefdef to\sppzttrra
\edef\spzzttrra{\expandafter\strip\spezttrra}
\immediate\write\fileftmp {\spzzttrra *.A}
\expandafter\xdef\csname\spzzttrra\endcsname{\sppzttrra}
\repeat \fi \immediate\closein\filefdef \fi}
\def\endfdef{\immediate\closeout\fileftmp   \ifnum\fdefcount>0
\immediate\openin\fileftmp \jobname.ftmp
\immediate\openout\filefdef \jobname.fdef
\immediate\write\filefdef {\the\fdefcount}   \indcount=0
\loop\ifnum\indcount<\fdefcount    \advance\indcount by 1
\immediate\read\fileftmp to\spezttrra
\edef\spzzttrra{\expandafter\strip\spezttrra}
\immediate\write\filefdef{\spzzttrra *.A}
\edef\spezttrra{\string{\csname\spzzttrra\endcsname\string}}
\iwritel\filefdef{\spezttrra}
\repeat  \immediate\closein\fileftmp \immediate\closeout\filefdef
\immediate\write16 {Panda Warning - fdef: Label(s) may have changed,
re-run TeX to get them right}\fi}
\def\iwritel#1#2{\newlinechar=-1
{\newlinechar=`\ \immediate\write#1{#2}}\newlinechar=-1}
\global\fdefcontre=0 \global\fdefcount=0 \global\indcount=0
%
%
\null
%
%
%
%

\input psfig

%
\loadamsmath
\loadeuler
\mathchardef\bbalpha="710B
\mathchardef\bbbeta="710C
\mathchardef\bbgamma="710D
\mathchardef\bbxi="7118
\mathchardef\bbomega="7121
\mathchardef\sdir="2D6E
\mathchardef\dirs="2D6F
\def\bal{{\bfmath\bbalpha}}
\def\bb{{\bfmath\bbbeta}}
\def\bga{{\bfmath\bbgamma}}

\def\ba{{\bfmath a}}
\def\bee{{\bfmath b}}
\def\be{{\bfmath e}}
\def\bq{{\bfmath q}}
\def\bg{{\bfmath g}}

\pageno=0\baselineskip=14pt
\nopagenumbers{
\line{\hfill SWAT/134}
\line{\hfill US-FT/38-96}
\line{\hfill\tt hep-th/9610142}
\line{\hfill October 1996}
\ifdoublepage \bjump\bjump\bjump\bjump\else\vfill\fi
\centerline{\capsone On the weak coupling spectrum of $N=2$}
\centerline{\capsone supersymmetric SU($n$) gauge theory}
\bjump\sjump
\centerline{\scaps  Christophe Fraser and Timothy J. Hollowood}
\sjump
\centerline{\sl Department of Physics, University of Wales Swansea,}
\centerline{\sl Singleton Park, Swansea SA2 8PP, U.K.}
\centerline{\tt  c.fraser t.hollowood @swansea.ac.uk}
\sjump
\bjump\bjump
\ifdoublepage
\vfill
\eject\null\vfill\fi
\centerline{\capsone ABSTRACT}\sjump
The weak coupling spectrum of BPS saturated states of
pure $N=2$ supersymmetric SU$(n)$ gauge theory is investigated. 
The method uses known results on the dyon spectrum of the
analogous theory with $N=4$ supersymmetry, along with the action on
these states of
the semi-classical monodromy transformations. For dyons whose magnetic
charge is not a simple root of the Lie algebra, it is found that the
weak coupling region is divided into a series of domains, for which
the dyons have different electric charge, separated by
walls on which the dyons decay. The proposed spectrum is shown to be
consistent with the exact solution of the theory at strong coupling in
the sense that the states at weak coupling can account for the
singularities at strong coupling. 
\sjump\vfill
\eject}
 \vfill

\yespagenumbers\pageno=1
%
%

\chapter{Introduction}

In [\Ref{SW1}] an exact expression for the prepotential of the effective action
of $N=2$ supersymmetric SU(2) gauge theory was found. 
Amongst other things, the prepotential determines the masses of any
BPS-saturated states in the theory, offering the hope of a complete
knowledge of the spectrum of such states. To be more specific,
the solution of Seiberg and Witten determines the mass of a BPS state, with
magnetic charge $g$ and electric charge $q$, as
$$
{\Bbb M}_{(g,q)}=\left\vert qa(u)+ga_{\rm D}(u)\right\vert
\nfr{MBPS}
where $a(u)$ and $a_{\rm D}(u)$ are functions of the gauge invariant
coordinate $u$ which parametrizes 
the moduli space of vacua. In particular, the exact
solution implies that there are points on the moduli space where a BPS
state becomes massless signalling that the original
effective action is no longer a good description of the long-range
physics. In fact there are two of these singularities which both appear at
strong coupling. They correspond to two arbitrary states,
one from each of the two sets
$(\pm1,2p)$ and $(\pm1,2p+1)$, $p\in{\Bbb Z}$, respectively, 
being massless. The
exact state in each set which becomes massless at the singularity, 
depends on the path taken to the singularity [\Ref{SW1}]. 

Because the functions $a(u)$ and $a_{\rm D}(u)$ are not single-valued
around a singularity, there exists
non-trivial monodromy on the moduli space. Taking a state $(g,q)$
around a singularity in the moduli space, one ends up with a state
$(g,q){\cal M}$, where ${\cal M}$ 
is a $2\times 2$ matrix acting by multiplication to the left. 
It would appear, therefore, that the monodromy transformations around
the two singularities, ${\cal M}^+$ and ${\cal M}^-$, should be
symmetries of the spectrum. However, this is not the case for reasons
which are explained below.

What was not determined in [\Ref{SW1}], is the actual spectrum of 
BPS states in the theory over the moduli space. 
At weak coupling, one can determine the
spectrum through a semi-classical analysis. This leads to a spectrum
of massive states consisting of towers of dyons with
charges $(\pm1,p)$, $p\in{\Bbb Z}$, as well as the gauge
bosons $(0,\pm1)$. States which appear at weak coupling
can then be continued into the region of strong coupling. There is
subtlety, however, arising from the fact that BPS states, 
which are generically below threshold for decay
into other states, can arrive at threshold for decay into other BPS
states on a Curve of Marginal Stability
(CMS). Upon crossing the curve, a BPS state can decay and disappear from the
spectrum. For SU(2), there is a single 
CMS is given by the one-dimensional curve on which 
$$
{\rm Im}\left({a_{\rm D}(u)\over a
(u)}\right)=0.
\efr
This curve is present only in the region of strong coupling and divides the
moduli space into two regions, one of which contains the weak coupling
regime [\Ref{F1},\Ref{AFS}]. The
curve passes through the two singularities on the moduli space.
The states which decay when crossing the CMS have been determined by a simple
argument in [\Ref{BF}]. It turns out that only two states, 
and their anti-particles, survive in the strong coupling region. 

The monodromy transformations ${\cal M}^\pm$ are not individually
symmetries of the BPS spectrum because paths encircling each of these
singularities pass through the CMS on which states can decay. 
However, a path that encircles both singularities need not pass
through the CMS and so the combined monodromy ${\cal M}^+{\cal M}^-$,
which is the semi-classical monodromy, or monodromy at infinity,
is a symmetry of the spectrum [\Ref{SW1}]. 

The analysis of Seiberg and Witten has been extended to theories with
an arbitrary gauge group: [\Ref{KLY},\Ref{AF},\Ref{KLT}] for SU($n$),
[\Ref{BL}] for SO($2n$), [\Ref{DS1}] for SO($2n+1$), [\Ref{AS}] for Sp($n$),
[\Ref{DS2}] for $G_2$ and [\Ref{AAG}] for all the other exceptional
cases. For larger gauge groups, the magnetic and 
electric charges are now rank$(g)$ vectors with respect to the
unbroken U$(1)^{{\rm rank}(g)}$ symmetry. There is an analogous
formula to \MBPS\ for the mass of BPS states
$$
{\Bbb M}_{(\bg,\bq)}
=\left\vert \bq\cdot\ba(u_j)+\bg\cdot\ba_{\rm D}(u_j)\right\vert,
\nfr{MGBPS}
where the vector quantities $\ba(u_j)$ and $\ba_{\rm D}(u_j)$ are determined
exactly in terms of the rank$(g)$ coordinates $u_j$ on the moduli
space. Just as in SU(2) case, there are singularities on the moduli
space and associated monodromies. However, it is now much more
complicated to determine the spectrum of BPS states, because, unlike
in SU(2), there are
many different CMS, depending upon the magnetic and electric charges
of the states involved,
some of which extend into the region of weak
coupling. Just as in
the SU(2) theory the existence of these CMS means that the
monodromies, but now even some of the semi-classical ones,
are not necessarily symmetries of the spectrum.

As a step towards a full understanding of the spectrum of BPS
states we consider the SU($n$) theory at weak coupling. In this limit,
we find the
positions of the CMS and propose, with reference to the spectrum of
the related $N=4$
supersymmetric theory, a form for the weak coupling spectrum of the
theory. We find that the weak coupling region is divided into various
domains by CMS, each domain having a different spectrum of BPS states.
We then show that the weak coupling spectrum is the minimal set
consistent with the exact
solution of the theory at strong coupling, in the sense that the
states which appear at weak coupling can account for the singularities
that appear at strong coupling. This suggests that like the SU(2)
theory, the weak coupling spectrum includes all the states that
appear at strong coupling.

\chapter{Semi-classical monodromies}

In this section we explain how monodromies appear on the moduli space
of vacua in the semi-classical regime. Our notation follows
closely that of [\Ref{BL},\Ref{DS1}].

In the semi-classical regime, we can parameterize the moduli space in
terms of the VEV of the classical Higgs field $\phi$. We can
use the freedom to perform
global gauge transformations to take $\phi$ to be in the
Cartan subalgebra of the Lie algebra of the gauge group, so
$\phi=\ba\cdot{\bfmath H}$, which defines the $n-1$ component vector $\ba$.
This does not entirely fix the global gauge
transformations since it leaves the freedom to perform discrete transformations
in the Weyl group. This discrete degree-of-freedom can be fixed, for
example, by demanding
that ${\rm Re}(\ba)$ lies in the fundamental Weyl chamber, i.e. 
with respect to some choice of simple roots $\bal_i$ of the Lie algebra
$$
{\rm Re}\left(\bal_i\cdot\ba\right)\geq0,\qquad i=1,2,\ldots,n-1.
\nfr{DSR} 
We will denote this region $W$.\note{The choice of the real part in
\DSR\ is not unique, it would be just as convenient to choose the
region ${\rm Re}(e^{-i\phi}\bal_i\cdot\ba)\geq0$, where $\phi$ is some
phase angle.}
The semi-classical regime is defined as the region for which
$\vert\bal_i\cdot\ba\vert\gg\Lambda$, where $\Lambda$ is the dynamically
generated mass scale in the theory, and it is only in this region that
$W$ coincides with the quantum moduli space.

The boundary of $W$ is made up of the walls $\partial W_i$ on which 
${\rm Re}(\bal_i\cdot\ba)=0$. 
The wall $\partial W_i$ is an invariant subspace under the action of the 
Weyl reflection in the simple root $\bal_i$, which we denote as $r_i$.
Moreover, a point $\ba\in\partial W_i$ is identified with
$r_i(\ba)$. The subspace $S_i$ of
$W$, of co-dimension two, defined by 
$\bal_i\cdot\ba=0$, is the $i^{\rm th}$ semi-classical singularity.
Obviously $S_i$ lies in the wall $\partial W_i$ and is left invariant 
by the Weyl reflection $r_i$. 

The prepotential  
can be calculated exactly in perturbation theory since it only
receives contributions at the tree and one-loop level.
This leads to the following expression for the quantity $\ba_{\rm
D}$ in the semi-classical regime
$$
\ba_{\rm D}={i\over2\pi}\sum_{\bal\in\Phi^+}\bal(\bal\cdot\ba)\left[\ln\left(
{\bal\cdot\ba\over\Lambda}\right)^2+1\right],
\nfr{AD}
where the sum is over $\Phi^+$ the set of positive roots of the Lie
algebra. The pre-potential also receives non-perturbative corrections,
however these do not affect the following weak coupling analysis.
The logarithm gives rise to singularities if $\bal\cdot\ba=0$, for some
root $\bal$ of the Lie algebra. These singularities occur precisely at the
points where classically one would expect the residual gauge symmetry to be
enhanced beyond the maximally abelian subgroup U$(1)^{n-1}$, due to
the fact that the gauge bosons associated to the root $\bal$, and its negative,
become massless. In the full quantum theory, these semi-classical singularities
are not in fact present, rather they are split into pairs of
singularities which coalesce in the limit $\Lambda\rightarrow0$. 
Furthermore, they are not caused by gauge bosons becoming
massless, which physically we would not expect because of strong
coupling effects,
rather they are a signal that certain magnetically charged dyons 
become massless. At weak coupling, however, these pairs of 
quantum singularities are not resolved and they give rise to the
effective semi-classical singularities which 
appear in \AD. We will have more to say about the quantum singularities in
section 5.

Notice that $\ba_{\rm D}$ is not a single-valued function of $\ba$ and hence
there are non-trivial monodromies as one encircles a singularity. The
origin of the monodromy is the fact that going around
the singularity $S_i$ involves identifying points on $\partial W_i$ by the 
Weyl reflection $r_i$. This is illustrated in figure 1.

\vbox{
\centerline{
\psfig{figure=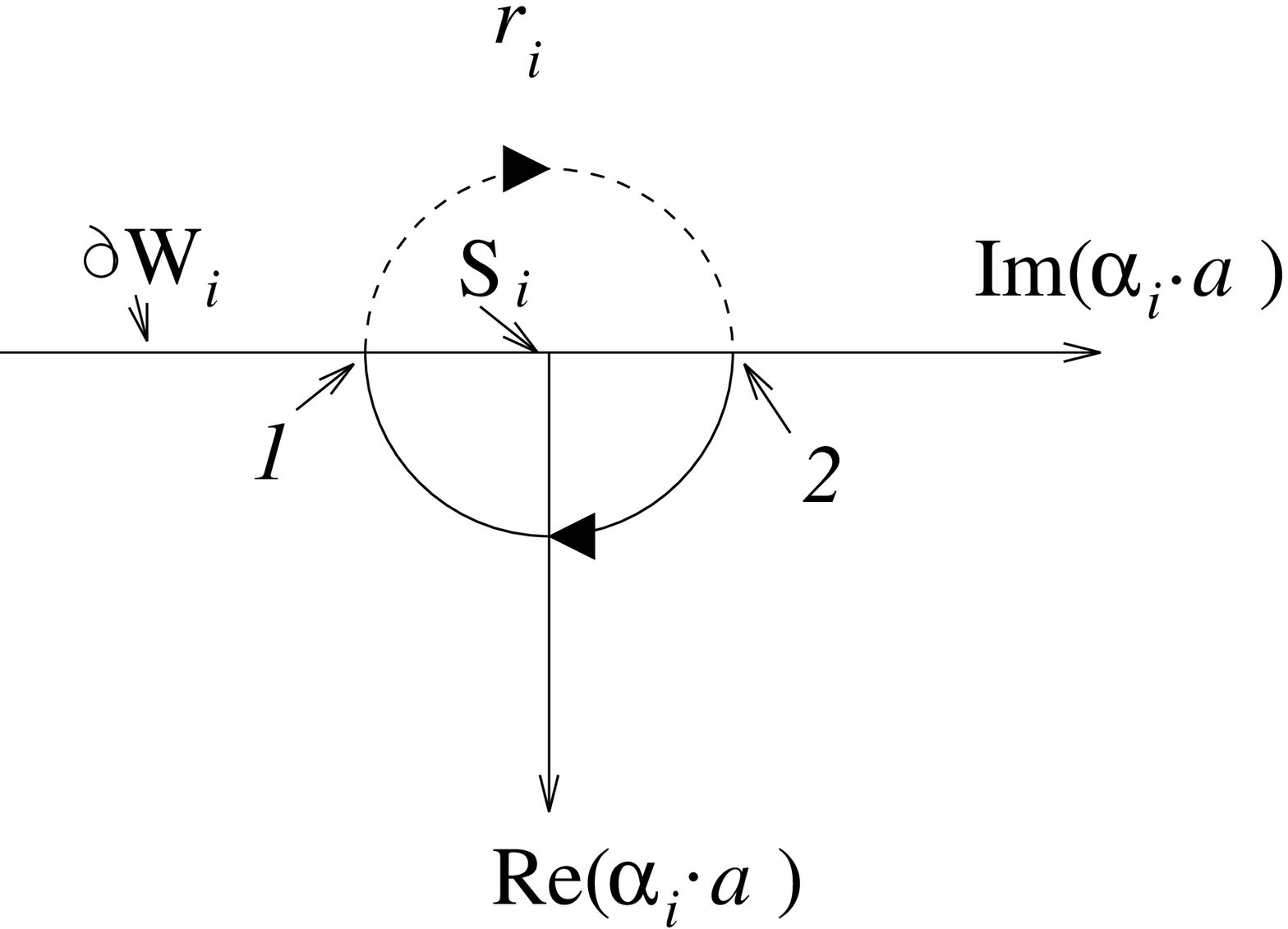,height=6cm}}
\bjump\bjump
\centerline{Figure 1. Path leading to monodromy $M_i$ around $S_i$.}
}
\bjump

Under this identification, 
the vector $(\ba_{\rm D},\ba)$ undergoes a monodromy transformation.
Defining a path around $S_i$ to be in a positive sense if on $\partial
W_i$ the Weyl reflection $r_i$ takes a point with ${\rm
Im}(\bal\cdot\ba)<0$, labelled as $1$ in figure 1, 
to a point with ${\rm Im}(\bal\cdot\ba)>0$, labelled as 2 in figure 1.
The monodromy can be derived by considering the
change of $\ba_{\rm D}$ along the path
$$
\ba(t)=\ba-\left(1-e^{i\pi t}\right)\bal_i(\bal_i\cdot\ba)/2,\qquad
0\leq t\leq1, 
\efr
giving
$$
\left({\ba_{\rm D}\atop\ba}\right)_2=
M_i\left({\ba_{\rm D}\atop\ba}\right)_1=\pmatrix{r_i&-\bal_i\otimes\bal_i\cr
0&r_i\cr}\left({\ba_{\rm D}\atop\ba}\right)_1.
\nfr{SCMO}
In a negative sense, the monodromy around $S_i$ is given by $M_i^{-1}$.
All other monodromies are generated by conjugation from the monodromies
$M_i^{\pm1}$ associated to the simple roots.

The existence of non-trivial monodromies implies that if we follow the
mass of a 
BPS state along some closed path in moduli space then we do not necessarily get
a state of the same mass at the end of the journey. If the path
encircles the singularity $S_i$ in a positive sense, starting with a
dyon of charge $(\bg,\bq)$ of mass ${\Bbb M}_{(\bg,\bq)}$, as we cross
through the wall of the Weyl chamber at point 1 to arrive at point 2, 
the quantities $(\ba_{\bf D},\ba)$ change discontinuously as in
\SCMO. Since the mass of a state must be continuous along a continous
path in moduli space, this implies that
the charges of the state must be transformed. From
$$
\left\vert(\bg,\bq)M_i\left({\ba_{\rm D}\atop\ba}\right)\right\vert=
{\Bbb M}_{(\bg,\bq)M_i},
\efr
it follows that we end up with a state 
of charge $(\bg,\bq)M_i$, where by definition $M_i$ acts by
matrix multiplication to the left.

\chapter{Curves of marginal stability at weak coupling}

To probe the region of weak coupling, we take $\Lambda\rightarrow0$
with $\bal_i\cdot\ba\neq0$. 
In this limit the mass of a state $(\bg,\bq)$ with
non-vanishing magnetic charge is dominated by the
magnetic charge: ${\Bbb M}_{(\bg,\bq)}\rightarrow C\vert\ba\cdot\bg\vert$. 
To see this notice that that as $\Lambda\rightarrow0$
$$
\ba_{\rm D}\rightarrow-\left\{{in\over\pi}\ln\Lambda\right\}\ba.
\efr

In section 4, we shall argue that dyons can only have a magnetic charge
which is a root of the Lie algebra. For such states, we will need to
know where the CMS are located. A dyon with magnetic charge 
$\bal+\bb$ is at threshold for decay to two dyons with magnetic charges
$\bal$ and $\bb$, where $\bal,\bb,\bal+\bb\in\Phi$, the root system of
$su(n)$, when
$$
C_{\bal,\bb}:\qquad\vert\ba\cdot(\bal+\bb)\vert=
\vert\ba\cdot\bal\vert+\vert\ba\cdot\bb\vert,\quad {\rm i.e.}\ 
\bal\cdot\ba/\bb\cdot\ba\in{\Bbb R}\geq0.
\nfr{TRI}
This defines the CMS that we denote $C_{\bal,\bb}$ which is a
hyperplane of co-dimension one in moduli space. Notice that in
the weak coupling limit the
CMS for states with non-vanishing magnetic charge in
the weak coupling limit, do not depend on the electric charge
of state. Obviously this will cease to be true 
away from the region of weak coupling.

The following properties of $C_{\bal,\bb}$ will be useful. (i) The curve only
has a non-trivial overlap with $W$ if $\bal$ and $\bb$ are either both positive
roots or both negative roots. (ii)  $C_{\bal,\bb}$ divides $W$ into two
distinct domains which we denote $D^\pm_{\bal,\bb}$. 

In order to demonstrate these two properties, consider the smooth map $x:\
W\rightarrow{\Bbb C}$ where
$$
x(\ba)={\ba\cdot\bal\over\ba\cdot\bb}.
\efr
Suppose, first of all, that
$\bal$ is a positive root and $\bb$ is a negative root.
Writing the real and imaginary parts $\ba\cdot\bal=a+ib$ and
$\ba\cdot\bb=p+iq$, then in $W$, by definition, $a\geq0$ and $p\leq0$. Consider
the subspace of $W$ which maps to $x(\ba)\in{\Bbb R}$. For real
$x(\ba)$ it follows that $x(\ba)=a/p\leq0$, so
that the image of
$W$ under the map can never be real and positive. 
From \TRI\ we see that
the CMS $C_{\bal,\bb}$ is
precisely the portion of $W$ on which $x(\ba)$ is real and positive. 
This demonstrates the first advertised property of 
$C_{\bal,\bb}$.

The second property follows immediately. Suppose that $\bal$ and $\bb$ are both
positive roots. Then under $x$, the image of $W$ can never be 
the negative real axis.
Furthermore the image of $C_{\bal,\bb}$ is precisely the positive real axis.
Hence $W$ is separated by $C_{\bal,\bb}$ into two distinct domains 
$D^+_{\bal,\bb}$ and
$D^-_{\bal,\bb}$ according to whether ${\rm Im}(x(\ba))>0$ or
${\rm Im}(x(\ba))<0$.  Obviously an
identical argument  can be followed if $\bal$ and $\bb$ are both
negative roots. 

Notice that $C_{\bal,\bb}$ passes through $\ba\cdot\bal=0$ and $\ba\cdot\bb=0$,
which are on the walls of $W$, corresponding to $x=0$ and
$x=\infty$, respectively. 

From this information it is possible to deduce all the CMS for a state with 
magnetic charge given by the root
$\bga$, which we suppose is a positive root. For each pair of positive
roots $\bal$ and $\bb$ such that $\bga=\bal+\bb$, then $C_{\bal,\bb}$ is a CMS
for the state with magnetic charge $\bga$. At this point
it will prove convenient to write the roots in terms of the weight
vectors of the $n$ dimensional representation of $su(n)$, $\be_i$,
$i=1,\ldots,n$, with inner products $\be_i\cdot\be_j=\delta_{ij}-(1/n)$. The
positive roots are then $\be_i-\be_j$, with $i<j$, and the simple roots are
$\bal_i=\be_i-\be_{i+1}$, $i=1,\ldots,n-1$.
So if $\bga=\be_i-\be_j$ then its CMS
are $C_{\be_i-\be_k,\be_k-\be_j}$ with $i<k<j$ corresponding to the
decay
$$
\left(\be_i-\be_j\right)\rightarrow\left(\be_i-\be_k\right)+
\left(\be_k-\be_j\right).
\efr
So for this state of magnetic
charge $\be_i-\be_j$, $W$ is divided into $2^{j-i-1}$ distinct regions
by its $j-i-1$ CMS given by the intersections 
$$
D_{\be_i-\be_{i+1},\be_{i+1}-\be_j}^{\epsilon_{i+1}}\cap\cdots\cap
D_{\be_i-\be_k,\be_k-\be_j}^{\epsilon_k}\cap\cdots\cap
D_{\be_i-\be_{j-1},\be_{j-1}-\be_j}^{\epsilon_{j-1}},
\efr
where $\epsilon_k=\pm$ for $i<k<j$. Notice that
there are no CMS for a state whose magnetic charge is a simple root.
Obviously an identical picture holds for the CMS of the negative roots. In fact
$C_{\bal,\bb}\equiv C_{-\bal,-\bb}$ and $D^\pm_{\bal,\bb}\equiv 
D^\pm_{-\bal,-\bb}$.

For SU(3) there is one positive root $\bal_1+\bal_2$ which is not
simple. So for 
the state of magnetic charge $\bal_1+\bal_2$, $W$ is divided into two regions
separated by the CMS $C_{\bal_1,\bal_2}$. Figure 2 shows a two-dimensional
cross-section of $W$. Notice, in this case, that $C_{\bal_1,\bal_2}$ 
intersects $\partial W_1$ in $S_1$ and $\partial W_2$ in $S_2$.

\vbox{
\centerline{
\psfig{figure=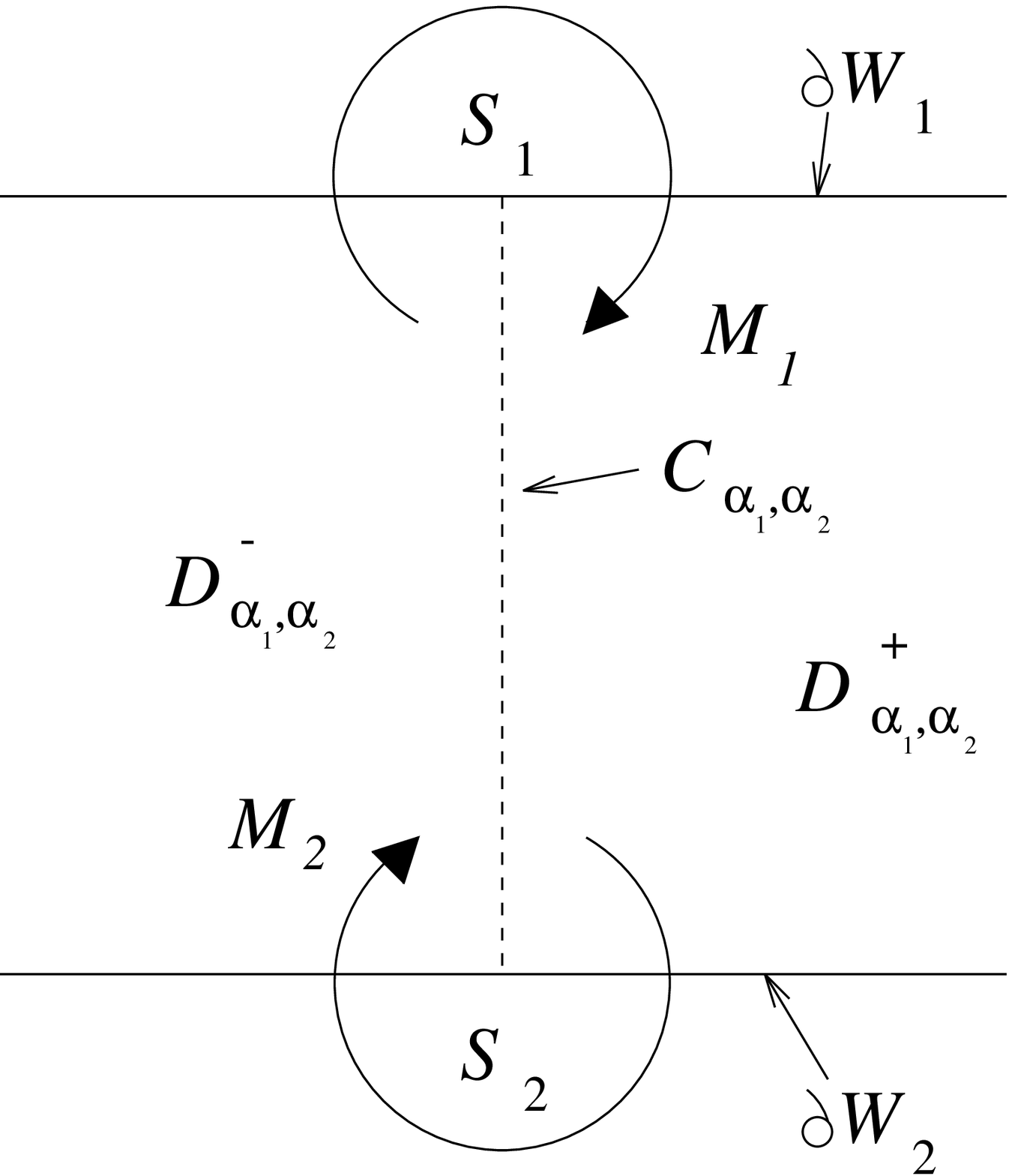,height=6cm}}
\bjump\bjump
\centerline{Figure 2. Two-dimensional cross-section of $W$ for SU(3).}
}
\bjump

\chapter{The weak coupling dyon spectrum} 

In this section, we use the knowledge of the dyon spectrum of the SU($n$) $N=4$
supersymmetric theory and the positions of the CMS to deduce the weak coupling
dyon spectrum of the $N=2$ theory. 

In the $N=4$ supersymmetric theory the Higgs field transforms as a
vector of the SO(6)
$R$-symmetry. The vacuum expectation value of the Higgs field  is therefore
specified by the 6 $(n-1)$-dimensional vectors $\ba^I$, $I=1,\ldots,6$. On a
special subspace of the moduli space for which 
$$
\ba^I=\xi^I\bee,
\nfr{SPEC} 
an analysis of the
semi-classical spectrum of monopole states has recently been performed
[\Ref{GL},\Ref{LWY}]. 
The monopole solutions are obtained by embedding the 't Hooft
Polyakov monopole of the SU(2) theory into the SU($n$) theory by specifying an
$su(2)$ subalgebra of $su(n)$ associated to a particular root of 
$su(n)$ [\Ref{W1}]. These solutions are therefore spherically symmetric.

The vector $\bee$ defines a set of simple roots via
$\bee\cdot\bal_i>0$.\note{We are assuming here that $\bee$ is not orthogonal to
any root so that the gauge symmetry is maximally broken to U$(1)^{n-1}$.} The
moduli space of a particular solution depends upon whether the root
$\bal$ is a simple root, or not. If the root is simple, then 
the moduli space is identical to that of the SU(2) monopole, i.e. it is of the
form
${\Bbb R}^3\times S^1$, corresponding to three translational and one charge
degrees of freedom. Semi-classical quantization proceeds exactly as in the
SU(2) theory and leads to a tower of dyons with magnetic charge
$\bal_i$ and electric charge $p\bal_i$, for $p\in{\Bbb Z}$. The
embedding of the anti-monopole gives a similar tower of dyons with opposite
magnetic charge.

For solutions corresponding to non-simple roots, the moduli space carries an
additional internal component ${\cal M}_0$
[\Ref{GL},\Ref{LWY},\Ref{MUR},\Ref{GIB1},\Ref{SC}]. 
In a certain asymptotic
limit, this internal part of the moduli space describes the fact 
that the solution for $\bal=\sum n_i\bal_i$ can be thought of a
superposition of $\sum n_i$ fundamental monopoles whose magentic
charges are simple
roots. The exact form of the moduli space is ${\Bbb R}^3\times({\Bbb
R}\times{\cal M}_0)/{\Bbb Z}$, where the ${\Bbb R}^3$ factor corresponds to the
translational degrees of freedom 
and the ${\Bbb R}$ factor to the overall charge
degree of freedom. The factor ${\cal M}_0$ in the asymptotic regime,
encodes the
relative positions and charge angles of the fundamental monopoles. On 
proceeding to a semi-classical quantization, it turns out
that there is unique harmonic form on
${\cal M}_0$ and hence there exists a bound-state of the
$\sum n_i$ fundamental monopoles at threshold carrying magnetic charge
$\bal$ [\Ref{GL},\Ref{GIB2}].
The resulting spectrum of states consists of a tower of
dyons with magnetic charge $\bal$ and electric charge $p\bal$, for $p\in{\Bbb
Z}$, where $\bal$ is any root of the algebra.

To summarize, 
on the special subspace \SPEC\ of the moduli space the spectrum of dyons,
whose magnetic charge is a root, is the same for all roots:
$$
N=4:\ \ \ \ \ (\bal,p\bal),\ \ \ \bal\in\Phi,\quad p\in{\Bbb Z}.
\efr
This result is a rather compelling piece of evidence for the old duality
conjecture of Goddard, Nuyts and Olive [\Ref{GNO}] in the context an $N=4$
supersymmetric gauge theory. 
Of course, one would like to know the full spectrum
of dyon states in the theory. In particular, requiring the theory to
have a full $S$-duality implies that there exist dyons with a
magnetic charge which are not roots of the algebra (but are in the root
lattice) [\Ref{DFH}], as in the SU(2) theory. To prove this
will require an analysis of the
appropriate multi-monopole moduli spaces. However, the $S$-duality conjecture
on the special subspace \SPEC, does not require the 
existence of dyon states whose electric and magnetic charges are not 
proportional.

The analysis of the 
$N=4$ theory has important implications for the weak coupling
spectrum of the
$N=2$ theory. For the case of SU(2), the complete spectrum of the 
$N=4$ supersymmetric theory consists of dyons with charges $(g,q)$, 
where $g$ and $q$ are co-prime integers. 
The states of magnetic charge $g$ can be
thought of as a bound-state of $g$ fundamental monopoles of unit magnetic
charge. The fact that such states appear in the spectrum of the $N=4$ theory,
is a consequence of the fact that there exists a bound-state of $g$
fundamental monopoles. In semi-classical quantization the bound-state
is manifested by the existence of a unique harmonic form
on the appropriate multi-monopole moduli space [\Ref{SEN},\Ref{POR}]. 
In the context of the $N=2$ theory, these bound-states would not exist
for the simple reason that they would require the existence of a
holomorphic harmonic form on the multi-monopole moduli space. But the existence
of such a form is ruled out because it would inevitably have an
anti-holomorphic partner, in contradiction to the uniqueness of the
harmonic form 
following from the analysis in the $N=4$ theory. Hence in the SU(2) theory the
weak coupling spectrum of dyons is only a subset of the states in the $N=4$
theory consisting of the dyons with unit magnetic charge, i.e. the states
$(\pm1,p)$, $p\in{\Bbb Z}$. We can think of these two towers of dyon
states as being associated to the single root of $su(2)$, and its negative.

Now consider the SU($n$) theory with $N=2$ supersymmetry. 
The analogue of the special subspace \SPEC\ is the subspace for which 
$$
\ba=e^{i\phi}\bee,
\nfr{SPC} 
for some phase $\phi$ and real vector $\bee$. 
Notice that the special subspace \SPC\ has dimension $n$ and
corresponds to the intersection of all the CMSs
$C_{\bal,\bb}$. Just as in the $N=4$ case we
define a set of simple roots with respect to $\bee$. Notice that these 
simple roots are the same simple roots as were defined
in \DSR\ with respect to ${\rm Re}(e^{-i\phi}\ba)$ (see the footnote
after \DSR). On the
special subspace \SPC\ we can determine the spectrum of dyons with magnetic
charges which are roots of the algebra by analogy with the preceding
argument for SU(2). As we have already remarked, in the $N=4$ theory the dyons
with magnetic charge which are not simple roots exist as bound states by virtue
of there being a unique harmonic form on the internal part of the 
multi-monopole moduli space ${\cal M}_0$. Since the harmonic form is unique, 
such bound-states will not exist in the
$N=2$ theory. Hence we expect that the spectrum of dyon states on the subspace
only consists of states with magnetic charge being a simple root, or its
negative:
$$
N=2:\ \ \ \ \ (\pm\bal_i,p\bal_i),\ \ \ p\in{\Bbb Z}.
\efr

We can immediately deduce from this analysis that, since in the weak
coupling regime there are no CMS for a state whose magnetic charge is a simple
root, the dyons $(\pm\bal_i,p\bal_i)$ must exist throughout the region
of weak coupling. A consistent picture of the weak coupling spectrum of
dyons can now be built up by transporting the basic dyons around
the semi-classical singularities.

As a preliminary to the general case, 
let us first consider SU(3). A two-dimensional cross-section
of $W$ appears in figure 2. Suppose we take the dyons states
$(\bal_1,p\bal_1)$ around $S_1$ in a positive sense. The states
are transformed by
the semi-classical monodromy transformation $M_1$, giving
$$
(\bal_1,p\bal_1)M_1=(-\bal_1,-(p+2)\bal_1).
\efr
These states are in the tower of anti-dyons associated to the
simple root $\bal_1$. Now consider taking the dyons $(\bal_1,p\bal_1)$ along
a path that winds around the singularity $S_2$
in a positive sense. The resulting states are 
$$
(\bal_1,p\bal_1)M_2=(\bal_1+\bal_2,p(\bal_1+\bal_2)+\bal_2).
\nfr{FS}
Similarly the same states taken along a path which encircles $S_2$ 
in a negative sense, end up as
$$
(\bal_1,p\bal_1)M_2^{-1}=(\bal_1+\bal_2,p(\bal_1+\bal_2)-\bal_2).
\nfr{SS}
One might now think that by taking the same states twice around $S_2$ one would
end up with more dyon states. For example
$$
(\bal_1,p\bal_1)M_2M_2=(\bal_1,p\bal_1-2\bal_2).
\efr
However, such states would be inconsistent with the fact that on the special
subspace \SPC\ we only expect dyons whose magnetic charge is a simple root to
have an electric charge which is proportional to the magnetic charge. The way
to resolve this problem, is to notice that in order 
to wind around the singularity
$S_2$ twice would necessarily entail crossing the CMS $C_{\bal_1,\bal_2}$. So
we have to assume that the states in \SS\ and \FS\ decay on crossing
$C_{\bal_1,\bal_2}$. This is consistent with the fact already noted, that on 
on the special subspace \SPC, which coincides with
$C_{\bal_1,\bal_2}$, there are no dyon
states whose magnetic charge is a non-simple root.

The two sets of states in \FS\ and \SS\ are therefore present in the disjoint 
regions of $W$, illustrated in figure 2, 
that we have defined in section 3 to be $D^\mp_{\bal_1,\bal_2}$,
respectively. Hence the
spectrum of dyon states at weak coupling can be summarized as follows. The dyon
states $(\bal_i,p\bal_i)$, $i=1,2$, 
are present throughout $W$. For the non-simple root
$\bal_1+\bal_2$, there is a tower of dyon states either \FS\ or \SS,
present in the two
regions $D^\mp_{\bal_1\bal_2}$, respectively, 
separated by $C_{\bal_1,\bal_2}$ on
which the states decay.  In each of these regions there is a
corresponding set of anti-dyons with opposite magnetic charge.

Suppose we had started with the dyons
$(\bal_2,p\bal_2)$, what states are created on taking these states around the
singularity $S_1$? In fact, no new states are created since
$$
(\bal_2,p\bal_2)M_1^{\pm1}=
(\bal_1,(p\pm1)\bal_1)M_2^{\mp1},
\efr
and furthermore the states $(\bal_2,p\bal_2)M_1^{\pm1}$ are present in
$D^\pm_{\bal_1,\bal_2}$, as they should for consistency.

Consider now the SU(4) case. It will be convenient to label the tower of dyons
associated to the simple roots as $Q_i=(\bal_i,p\bal_i)$. Just as in the
SU(3) case, starting from the dyons $Q_1$, the dyons $Q_1M_2^{\pm1}$,
of magnetic charge $\bal_1+\bal_2$, will be
generated in the disjoint regions
$D^\mp_{\bal_1,\bal_2}$, by winding around $S_2$ in either a positive or
negative sense. These states are identical to $Q_2M_1^{\mp1}$.
In a similar way, dyons $Q_2M_3^{\pm1}$, of magnetic charge
$\bal_2+\bal_3$, will be generated in
the disjoint regions $D^\mp_{\bal_2,\bal_3}$, by winding around the singularity
$S_3$ in either a positive or negative sense. These states are identical to
$Q_3M_2^{\mp1}$.
 
Now consider the affect of taking the states $Q_1M_2^{\pm1}$
around the singularity $S_3$. A two-dimensional cross-section of the wall
$\partial W_3$ is illustrated in figure 3 showing how the regions
$D^\pm_{\bal_1,\bal_2}$ and the CMS $C_{\bal_1,\bal_2}$ intersect it.

\vbox{
\centerline{
\psfig{figure=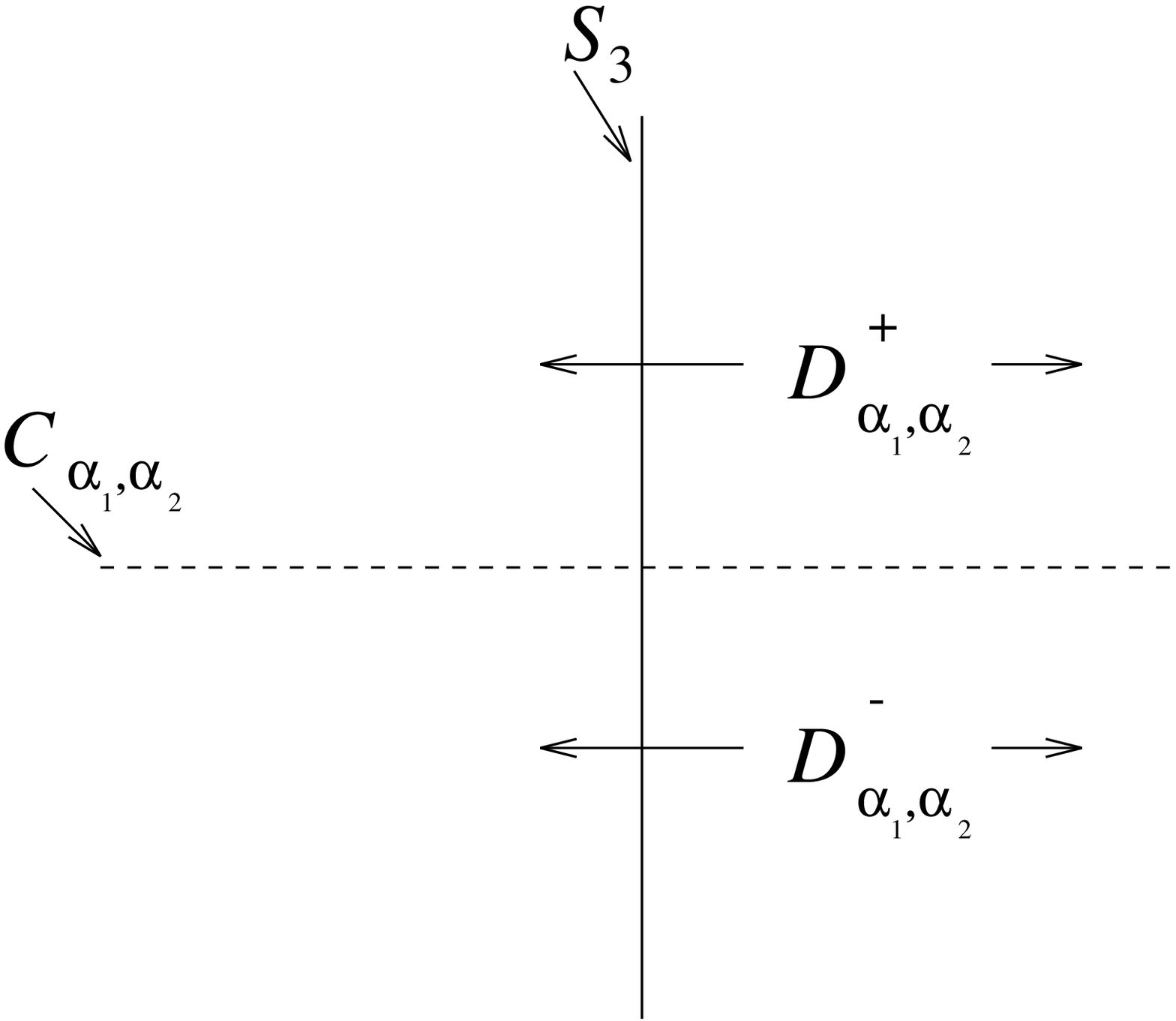,height=6cm}}
\bjump\bjump
\centerline{Figure 3. Cross-section of the wall $\partial W_3$ showing
the two}
\centerline{distinct regions for dyons of magnetic charge $\bal_1+\bal_2$.}
}
\bjump
 
Suppose we take the states $Q_1M_2^{\pm1}$ around the singularity
$S_3$ in a positive or negative sense. 
We end up with four regions each containing
a different set of states:
$$\eqalign{
D^-_{\bal_1,\bal_2+\bal_3}\cap D^-_{\bal_1+\bal_2,\bal_3}:&\qquad
Q_1M_2M_3\cr
D^-_{\bal_1,\bal_2+\bal_3}\cap D^+_{\bal_1+\bal_2,\bal_3}:&\qquad
Q_1M_2M_3^{-1}\cr
D^+_{\bal_1,\bal_2+\bal_3}\cap D^-_{\bal_1+\bal_2,\bal_3}:&\qquad
Q_1M_2^{-1}M_3\cr
D^+_{\bal_1,\bal_2+\bal_3}\cap D^+_{\bal_1+\bal_2,\bal_3}:&\qquad
Q_1M_2^{-1}M_3^{-1}.\cr}
\efr 
This is illustrated in figure 4. 

\vbox{
\centerline{
\psfig{figure=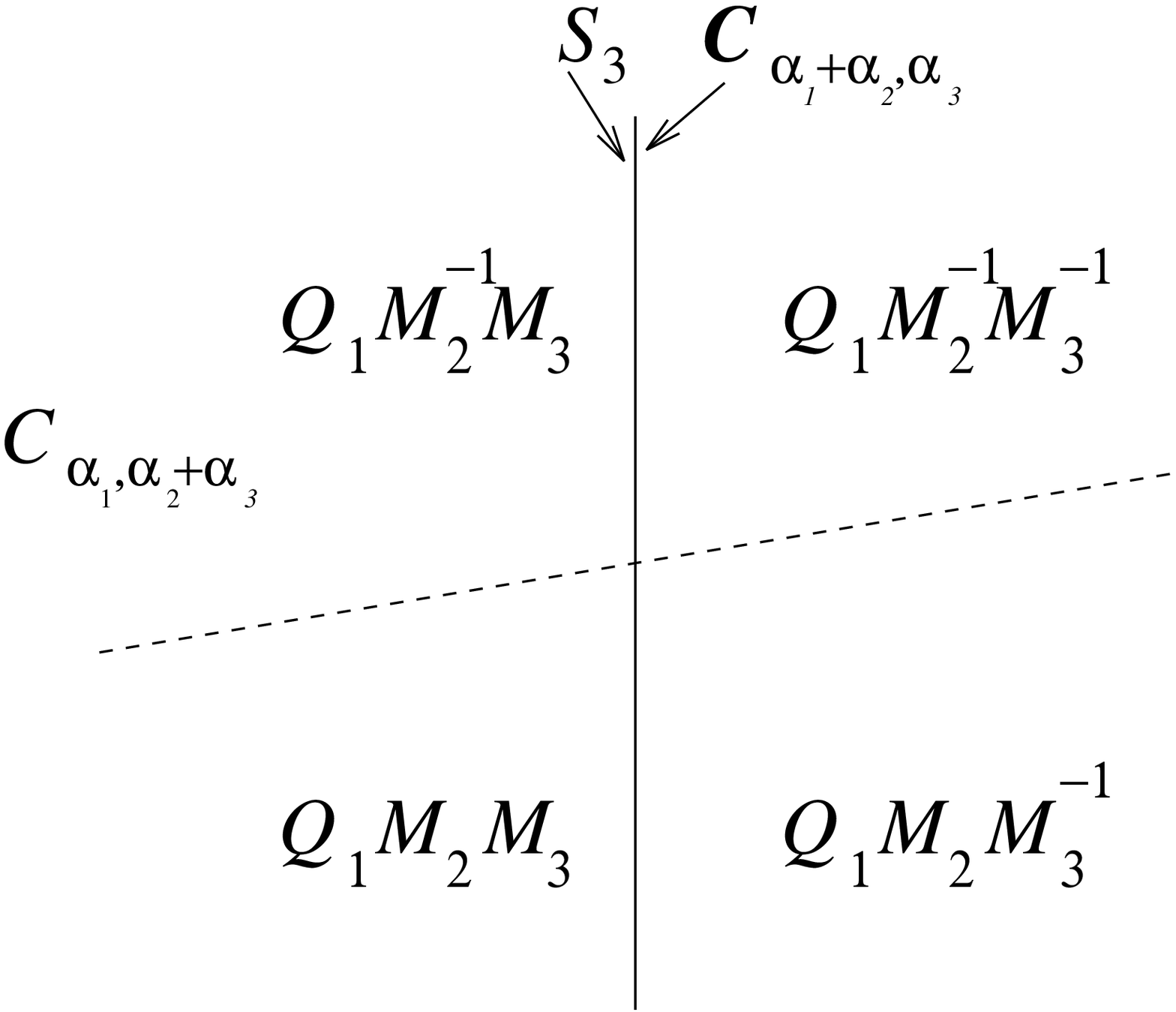,height=6cm}}
\bjump\bjump
\centerline{Figure 4. Cross-section of the wall $\partial W_3$ showing
the four}
\centerline{distinct regions for dyons of magnetic charge
$\bal_1+\bal_2+\bal_3$.}
}
\bjump

Notice that under the Weyl reflection
$r_3$ the CMS $C_{\bal_1,\bal_2}$ is mapped into the CMS
$C_{\bal_1,\bal_2+\bal_3}$. Moreover, the CMS $C_{\bal_1+\bal_2,\bal_3}$
intersects $\partial W_3$ in $S_3$. No new states are generated by
considering other possibilities, for example $Q_1M_2M_3\simeq
Q_2M_3M_1^{-1}\simeq  Q_3M_2^{-1}M_1^{-1}$.

The generalization to SU($n$) follows in an obvious way. We can generate a
complete set of states by starting with the dyons $Q_i$, whose
magnetic charges are simple roots. First of all, $Q_iM^{\pm1}_i$
generates the tower of anti-dyons $(-\bal_i,-(p\pm2)\bal_i)$. 
Dyons with magnetic
charges which are not simple roots are generated by
$$
Q_iM_{i+1}^{\epsilon_{i+1}}M_{i+2}^{\epsilon_{i+2}}\cdots
M_{j-1}^{\epsilon_{j-1}},
\nfr{DSET}
where $\epsilon_k=\pm1$ and we take $j>i+1$. Using the fact that
$\bal_i+\cdots+\bal_{j-1}=\be_i-\be_j$, these dyons have
charges
$$
\left(\be_i-\be_j,p'(\be_i-\be_j)-\sum_{k=i+1}^{j-1}\epsilon_k
\left(\be_i-\be_k\right)\right),
\nfr{DCH}
for some integer $p'$ related to $p$.
Each of the $2^{j-i-1}$ sets in \DSET\ is
present in a different region of $W$ given by
$$
D_{\be_i-\be_{i+1},\be_{i+1}-\be_j}^{-\epsilon_{i+1}}\cap\cdots\cap
D_{\be_i-\be_k,\be_k-\be_j}^{-\epsilon_k}\cap\cdots\cap
D_{\be_i-\be_{j-1},\be_{j-1}-\be_j}^{-\epsilon_{j-1}}.
\nfr{DRE}
Each of these regions is separated by a CMS on which the dyons, which
all have magnetic charge $\be_i-\be_j$, decay.

It is straightforward to see that the same set of dyons can be generated by
starting with $Q_k$, where $i\leq k\leq j$. In fact the sets
$$
Q_k\left(M_{k+1}^{\epsilon_{k+1}}\cdots M_{j-1}^{\epsilon_{j-1}}\right)
\left(M_{k-1}^{-\epsilon_{k-1}}\cdots M_{i+1}^{-\epsilon_{i+1}}\right),
\nfr{MDC}
coincide exactly with those in \DSET.

In the next section we shall require the semi-classical monodromy
associated to a tower of dyon states. If the tower of dyons 
have magnetic charge $\bal$, this is defined to be the
monodromy around the singularity $\bal\cdot\ba=0$.
So for the tower of dyons 
associated to the simple roots $Q_i$ the semi-classical 
monodromy is simply $M_i$.
For dyons whose magnetic charges are not simple roots, we can find the
monodromy by writing the tower of states as in \MDC, i.e. as 
$Q_iM$, where $M$ is some product of monodromies of the simple roots.
In order to encircle the semi-classical singularity for these dyons, we 
must first of all follow a path with monodromy $M^{-1}$, which
transforms the states to $Q_i$, i.e. with magnetic charge $\bal_i$.
Then we loop the singularity $S_i$ with monodromy $M_i$ and
finally follow a path with monodromy $M$ to return to the starting
point. So the total semi-classical monodromy for the dyons $Q_iM$ is 
$$
M^{-1}M_iM.
\nfr{SCT}
Since a given tower can be formed in different ways \MDC, there follow a
series of identities. For example $Q_1M_2\simeq Q_2M_1^{-1}$ implies that
$$
M_2^{-1}M_1M_2=M_1M_2M_1^{-1}.
\efr
Such identities can be proved by explicit computation.

There exists a symmetry of the spectrum which 
follows from the fact that the theory has a discrete global symmetry ${\Bbb
Z}_{2n}$ on the moduli space, being the discrete remnant
left over from the the classical U$(1)_{\cal R}$ symmetry which is
broken by quantum effects. This symmetry acts as
$$
u_j\rightarrow e^{i\pi j/n}u_j.
\efr
In the semi-classical regime this means $\ba\rightarrow
e^{i\pi/n}\ba$ and it is a simple matter to work out the associated
transformation on the spectrum at weak coupling. First of all,
$$
\left({\ba_{\rm D}\atop\ba}\right)\rightarrow
e^{i\pi/n}\pmatrix{1&-1\cr 0&1\cr}\left({\ba_{\rm D}\atop\ba}\right),
\efr
so that on the states the symmetry acts as
$$
(\bg,\bq)\rightarrow(\bg,\bq)\pmatrix{1&1\cr 0&1\cr}.
\efr
Since this transformation relates dyon states in the same
tower, it is clearly a symmetry of the weak coupling
spectrum that we have proposed.

\chapter{Strong coupling singularities}

In this section, we consider the relation between our proposed weak
coupling spectrum and the behaviour of the theory at strong coupling.
Underlying the exact solution of the model is a hyperelliptic
curve [\Ref{SW1},\Ref{KLY},\Ref{AF}] 
which appears in the following way. In order to parameterize the
whole moduli space it is necessary to introduce gauge invariant 
coordinates $u_j$, $j=2,3,\ldots,n$. At weak coupling, these are related to the
variables $\ba$ in the following way:
$$
\prod_{i=1}^n\left(x-\be_i\cdot\ba\right)=x^n-\sum_{j=2}^n u_jx^{n-j}
\equiv{\cal W}_n(x,u_j),
\efr
where $x$ is some auxiliary variable and we have defined the
polynomial ${\cal W}_n(x,u_j)$ in $x$. The quantum moduli space of the
SU($n$) theory is then precisely the moduli space of the curve
$\cal C$ defined by:
$$
{\cal C}:\qquad y^2=\left({\cal W}_n(x,u_j)\right)^2-\Lambda^{2n}.
\efr
Notice that the polynomial on the right-hand-side 
factors into ${\cal W}_n(x,u_j)\pm\Lambda^n$.
Let us suppose that the zeros of ${\cal W}_n(x,u_j)$ are located at
$z_i(u_j)$, $i=1,\ldots,n$. 
It follows that the zeros of the right-hand-side are located at
$$
z_i^\pm(u_j,\Lambda)\equiv z_i(u_2,\ldots,u_{n-1},u_n\pm\Lambda^n).
\efr
The curve $\cal C$ can be represented as the two-sheeted $x$
plane with cuts running between pairs $z^+_i$ and $z^-_i$. 

It will be convenient to consider a two-dimensional slice of moduli
space defined by $u_j=0$, $j=1,\ldots,n-2$, and $v\equiv u_n$, a fixed
number which we take to be real and greater than $\Lambda^n$. 
The slice is parameterized by $u\equiv u_{n-1}$. The slice
cuts across $2n$ singular points $S_j^\pm$  given by the $n$ solutions
of each of the two equations:
$$
\left({-u\over n}\right)^n+\left({v\pm\Lambda^n\over n-1}\right)^{(n-1)}=0,
\efr
for fixed $v$. The slice is illustrated in figure 5.

\vbox{
\centerline{
\psfig{figure=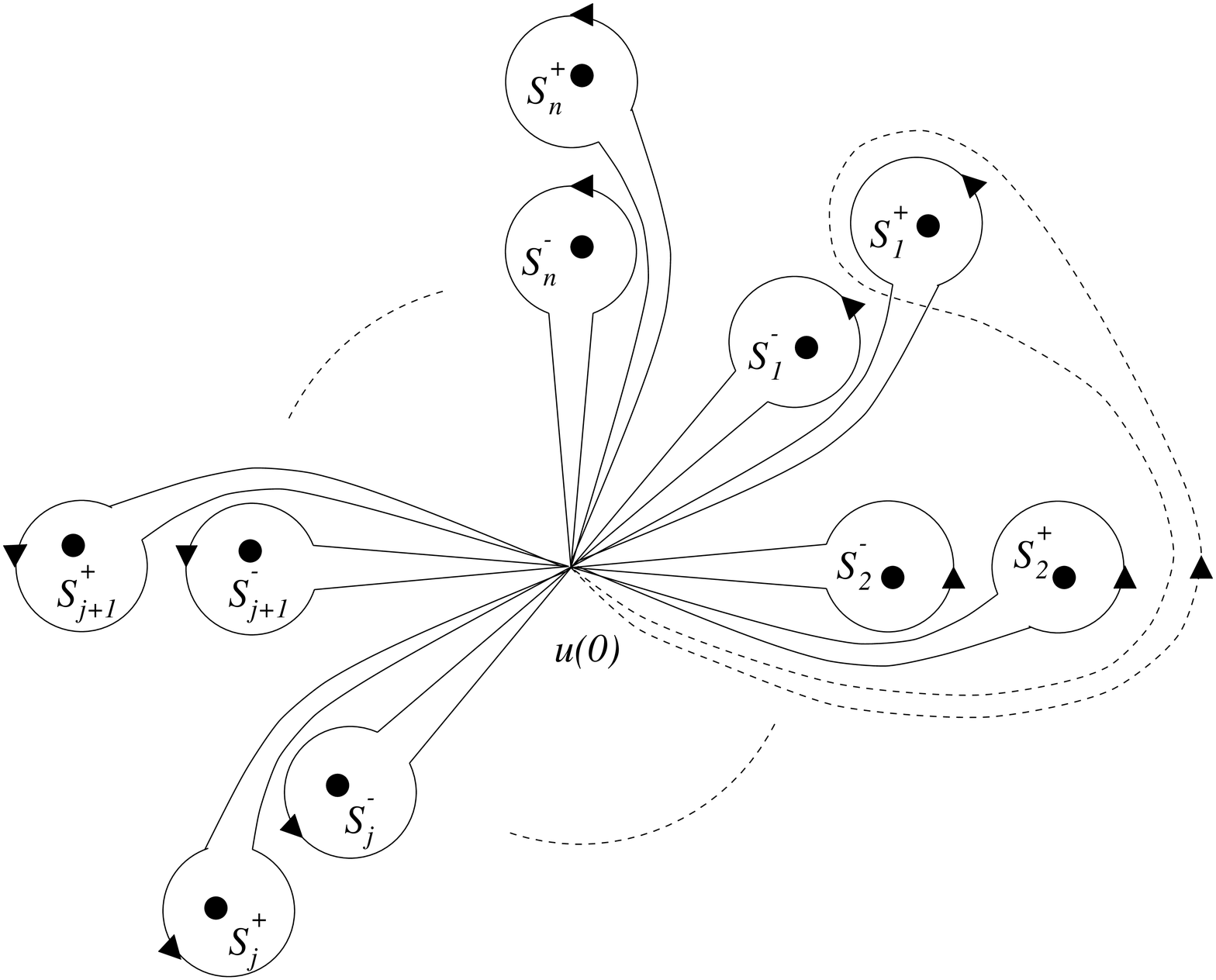,height=6cm}}
\bjump\bjump
\centerline{Figure 5. Singularities and paths on the slice
$u_j=(0,\ldots,0,u,v)$, for fixed $v$.}
}
\bjump

At the origin, labelled as $u(0)$ in figure 5, 
the polynomial is simply ${\cal W}_n(x)=x^n-v$, so that the
branch points of ${\cal C}$ are arranged in pairs at each $n^{\rm th}$
root of unity:
$$
z_j^\pm=\omega^{j-1}\left(v\pm\Lambda^n\right)^{1/n},
\efr
where $\omega=e^{2\pi i/n}$.
The positions of the branch points are illustrated in figure 6 which also 
shows a set of homology cycles $\gamma_j$, $\delta_j$, with $j=1,\cdots,n$.
In the figure only the parts of the cycles on the upper sheet are shown.

\vbox{
\centerline{
\psfig{figure=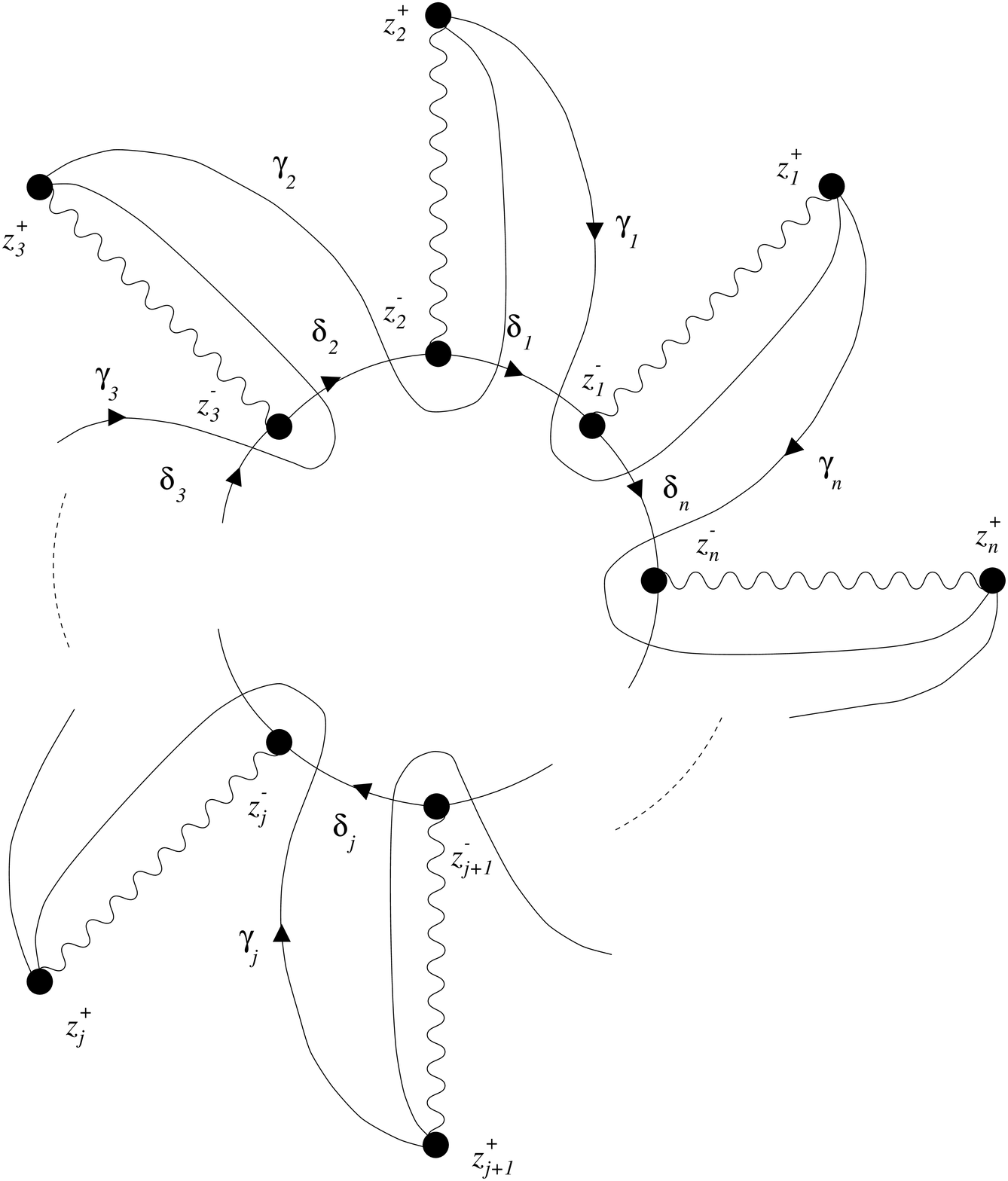,height=6cm}}
\bjump\bjump
\centerline{Figure 6. Basis of homology cycles of $\cal C$.}
}
\bjump

A singularity corresponds to a situation when two zeros, either
$z_i^+$ with $z_j^+$, or $z_i^-$ with $z_j^-$, coalesce, which can also
be described as the vanishing of a certain homology
cycle of $\cal C$. By carefully tracing out the paths figure 5,
one finds that the singularities $S_j^\pm$ correspond to following 
vanishing cycles:
$$\eqalign{
&S_j^+\qquad {\rm i.e.}\qquad z_j^+\sim z_{j+1}^+\qquad {\rm with}
\qquad\gamma_j\cr
&S_j^-\qquad {\rm i.e.}\qquad z_j^-\sim z_{j+1}^-\qquad {\rm with}
\qquad\delta_j.\cr}
\efr
The other possible singularities can be reached by taking more
complicated paths, for example the dotted path in figure 5
corresponds to $z_1^+$ coalescing with $z_3^+$ along the vanishing
cycle $\gamma_1+\gamma_2$.

In the semi-classical regime, the origin of the slice, $u(0)=(0,\ldots,0,v)$,
corresponds to
$$
\be_{\sigma(j)}\cdot\ba=v^{1/n}\omega^{j-1}, 
\nfr{SSS}
where $\sigma(j)$ is the following permutation of $\{1,2,\ldots,n\}$:
$$
\sigma(j)=\cases{2j-1&$j\leq\left[{n+1\over2}\right]$\cr
2(n-j+1)&$j>\left[{n+1\over2}\right]$\cr}.
\efr
(In the above $[x]$ is the integer part of $x$.)
The purpose of the permutation is to ensure that $u(0)$ lies in the
fundamental Weyl chamber $W$ according to \DSR\ which requires ${\rm
Re}(\be_i\cdot\ba)\geq{\rm Re}(\be_j\cdot\ba)$, for $i<j$. 
In the weak coupling limit
$\Lambda\rightarrow0$, each of the pairs of singularities $S_j^\pm$
coalesce to become the semi-classical singularity 
$$
\left(\be_{\sigma(j)}-\be_{\sigma(j+1)}\right)\cdot\ba=0.
\efr
It is straightforward to show in this limit that $u(0)$ lies in the region 
$$
\left(D_{\bal_1,\bal_2}^+\cap
D_{\bal_3,\bal_4}^+\cap\cdots\right)\cap
\left(D_{\bal_2,\bal_3}^-\cap
D_{\bal_4,\bal_5}^-\cap\cdots\right).
\nfr{REG}
One also finds through a numerical analysis that,
in the weak coupling limit, the 
paths to the singularities from $u(0)$, 
given by the vanishing of the cycles $\gamma_j$ and $\delta_j$, 
do not cross any CMS.

There is a concrete relation between the curve ${\cal C}$ and the quantities
$\ba(u_j)$ and $\ba_{\rm D}(u_j)$, appearing in the mass formula of BPS
states, provided by a very particular one-form  
$\lambda$ [\Ref{AF},\Ref{KLT}]:
$$
\bg_\eta\cdot\ba_{\rm D}+\bq_\eta\cdot\ba=\oint_{\eta}\lambda,
\nfr{RELA}
where $\eta$ is some homology cycle on ${\cal C}$. This
implies a mapping between homology cycles $\eta$ and charges 
$(\bg_\eta,\bq_\eta)$ whose explicit form we shall write down below.
As explained in [\Ref{KLT}], the 
intersection number of two cycles is given by
$$
\eta\cap\rho=\bg_\eta\cdot\bq_\rho-\bq_\eta\cdot\bg_\rho.
\nfr{RCS}
At the singularity described by the 
vanishing of a cycle $\eta$, the dyons of charge
$\pm(\bg_\eta,\bq_\eta)$ become massless since $\oint_\eta\lambda=
\bg_\eta\cdot\ba_{\rm D}+\bq_\eta\cdot\ba=0$.
In fact, it is the existence of these massless states that causes the 
breakdown of the effective action. 
Furthermore, if a singularity corresponds to the states $\pm(\bg,\bq)$
becoming 
massless, then the monodromy around the singularity can be calculated 
using perturbation theory in dual variables. On finds [\Ref{KLT}]
$$
{\cal M}_{(\bg,\bq)}=\pmatrix{1-\bq\otimes\bg&-\bq\otimes\bq\cr
\bg\otimes\bg&1+\bg\otimes\bq\cr}=1+\left({-\bq\atop\bg}\right)\otimes
(\bg,\bq),
\nfr{SCM}
with the property that
$$
(\bg,\bq){\cal M}_{(\bg,\bq)}=(\bg,\bq).
\efr

We now determine the mapping between vanishing cycles and dyon
charges. First of all, the vanishing cycles $\gamma_j$ and
$\delta_j$ correspond to massless dyons of magnetic charge 
$\pm(\be_{\sigma(j)}-\be_{\sigma(j+1)})$. 
Secondly, the 
cycles $\gamma_j$ and $\delta_j$ have the following intersection numbers
$$\eqalign{
&\gamma_j\cap\gamma_{j+1}=\delta_j\cap\delta_{j+1}=1,\cr
&\gamma_j\cap\delta_j=2,\quad\gamma_j\cap\delta_{j-1}=-2,\quad
\gamma_j\cap\delta_{j+1}=0.\cr}
\nfr{INTC}
The vanishing of the cycles $\gamma_j$ correspond to dyons of charge
$$\eqalign{
\left(\bal_{\sigma(j)}+\bal_{\sigma(j)+1},\bal_{\sigma(j)+1}\right),&\qquad
1\leq j\leq\left[{n-1\over2}\right]\cr
(-)^n\left(\bal_{n-1},\half(1-(-)^n)\bal_{n-1}\right),&\qquad j=
\left[{n+1\over2}\right]\cr
-\left(\bal_{\sigma(j)-2}+\bal_{\sigma(j)-1},\bal_{\sigma(j)-2}\right),&\qquad
\left[{n+3\over2}\right]\leq j\leq n-1\cr
-(\bal_1,0),&\qquad j=n,\cr}
\nfr{ASC}
along with
$$
\left(\bg_{\delta_j},\bq_{\delta_j}\right)=
\left(\bg_{\gamma_j},\bq_{\gamma_j}+\bg_{\gamma_j}\right).
\efr
This assignment is the unique choice consistent with \RCS\ and \INTC, up to the
freedom to perform the overall transformation
$$
(\bg,\bq)\rightarrow(\bg,\bq)\pmatrix{1&p\cr 0&1\cr},\qquad p\in{\Bbb Z}.
\efr

A more revealing way to write the charges of the dyons in \ASC, whose magnetic
charges are not simple roots, is
$$\eqalign{
(\bal_{\sigma(j)},0)M_{\sigma(j)+1}, &\qquad
1\leq j\leq\left[{n-1\over2}\right]\cr
-(\bal_{\sigma(j)-2},\bal_{\sigma(j)-2})M_{\sigma(j)-1}^{-1}, &\qquad
\left[{n+3\over2}\right]\leq j\leq n-1.\cr}
\nfr{DBM}
Recall, the electric charges of
dyons whose magnetic charges are not simple roots, is different in
each of the domains in \DRE. Furthermore, as we have already remarked,
the paths to the singularities along the vanishing cycles $\gamma_j$
and $\delta_j$, in the weak coupling limit, do not pass through any
CMS, therefore the singularities $S_j^\pm$ must be caused by 
dyons which are {\it actually\/} in the spectrum at $u(0)$.
From \REG, we see that the dyons with charges \DBM\ are indeed 
present at $u(0)$.
It is an important and highly non-trivial check of our proposal
for the weak coupling spectrum that the dyons \DBM, which are responsible
for the singularities $S_j^\pm$ in the slice, are actually present in the
spectrum at $u(0)$. 

The monodromies around the singularities follow from the charges of
the states and the formula \SCM. However, they can also be deduced directly
from the curve $\cal C$. For example, a loop around the singularity
at which $\gamma_i$ vanishes, makes $z_i^+$ interchange with $z_{i+1}^+$.
The way that the homology cycles are transformed can be determined
from the Picard-Lefshetz formula [\Ref{AGV}]. This states that the
action of the monodromy around a singularity corresponding to a
vanishing cycle $\nu$ on a cycle $\eta$ is
$$
{\cal M}_\nu(\eta)=\eta-(\eta\cap\nu)\eta.
\nfr{PL}
It is easy to verify that the monodromy computed from \PL\ 
is consistent with that computed from \SCM\ by virtue of the relation
$$
(\bg_{{\cal M}_\nu(\eta)},\bq_{{\cal M}_\nu(\eta)})
=(\bg_\eta,\bq_\eta){\cal M}_{(\bg_\nu,\bq_\nu)},
\efr
and \RCS.

As a non-trivial 
check of these assignments \ASC\ we now show that they lead to the correct 
semi-classical monodromies in \SCMO. In the full quantum theory, the 
semi-classical singularities split into two singularities. 
With reference to figure 5, we see that the product of the two
monodromies corresponding to the vanishing of the 
cycles $\gamma_j$ and $\delta_j$, taken in that order, should give the
appropriate semi-classical monodromy. So if the pair of quantum
singularities correspond to the massless dyons
$(\bal_j,p\bal_j)M$ and $(\bal_j,(p+1)\bal_j)M$, respectively,
where $M$ is some semi-classical monodromy
transformation and $p$ is some integer, then,
taking into account that monodromy transformations act on charge
vectors to the left, we should have
$$
M^{-1}M_jM={\cal M}_{(\bal_j,p\bal_j)M}{\cal M}_{(\bal_j,(p+1)\bal_j)M},
\nfr{NREL}
where the left-hand-side is the semi-classical monodromy associated to
the tower of dyons $Q_jM$ in \SCT. To verify \NREL, firstly one shows
$$
M_j={\cal M}_{(\bal_j,p\bal_j)}{\cal M}_{(\bal_j,(p+1)\bal_j)},
\nfr{EO}
which follows from the explicit expressions in \SCMO\ and \SCM. Then
from \SCM\ it is easy to see that for any semi-classical monodromy $M$
$$
{\cal M}_{(\bg,\bq)M}=M^{-1}{\cal M}_{(\bg,\bq)}M.
\nfr{ET}
Both \EO\ and \ET\ together imply \NREL.

\chapter{Final comments}

The problem of finding the spectrum of BPS saturated states in these
theories is complicated by the existence of CMS. At weak coupling we
have found the CMS, but in order to make progress across the whole of the
moduli space it would be desirable to have a 
description of the regions where particular dyon states are stable
in terms of the auxiliary
hyper-elliptic curve ${\cal C}$ which underpins the exact solution of
the theory. Some recent progress in this
direction has been made in [\Ref{KLMVW}]. 

Our more modest aim has been
to propose a form for the semi-classical spectrum of
BPS saturated states in $N=2$ supersymmetric SU$(n)$ gauge theory at
weak coupling based on the known spectrum of the associated $N=4$
supersymmetric theory and the positions of the CMS. 
The spectrum was shown to be consistent with the exact
solution of the theory at strong coupling, however, as we have
mentioned, the dyon spectrum at strong coupling is still an open question. In
particular, one would like to known the geometry of the CMS at strong
coupling. 

Lastly, it would be an interesting challenge to understand the picture
of the spectrum at weak coupling directly in terms of semi-classical
quantization. In fact it is possible to show directly how monopoles decay on a
CMS at weak coupling, since on these subspace the moduli space of a
monopole changes discontinously.

\bjump
We would like to thank Marco Kneipp for useful conversations. 
TJH is supported by a PPARC Advanced Fellowship and would like to
thank the Department of Particle Physics at the Universidad de Santiago de
Compostela, where most of this work was done, for their hospitality
and IBERDROLA for assistance under their Science and Technology Programme.

\references

\beginref
\Rref{W1}{E. Weinberg, Nucl. Phys. {\bf B167} (1980) 500}
\Rref{SW1}{N. Seiberg and E. Witten, Nucl. Phys. {\bf B426} (1994) 19,
{\tt hep-th/9407087}}
\Rref{AF}{P.C. Argyres and A.E. Faraggi, Phys. Rev. Lett. {\bf74}
(1995) 3931, {\tt hep-th/9411057}}
\Rref{DS1}{U.H. Danielsson and B. Sundborg, Phys. Lett. {\bf B358}
(1995) 273, {\tt hep-th/9504102}}
\Rref{BL}{A. Brandhuber and K. Landsteiner, Phys. Lett. {\bf B358}
(1995) 73, {\tt hep-th/9507008}}
\Rref{KLT}{A. Klemm, W. Lerche and S. Theisen, Int. J. Mod. Phys. {\bf
A11} (1996) 1929, {\tt hep-th/9505150}}
\Rref{KLY}{A. Klemm, W.Lerche and S. Yankielowicz, Phys. Lett. {\bf
B344} (1995) 169, {\tt hep-th/9411048}}
\Rref{KLMVW}{A. Klemm, W. Lerche, P. Mayr, C. Vafa and N. Warner, 
{\tt hep-th/9604034}}
\Rref{BF}{A. Bilal and F. Ferrari, Nucl. Phys. {\bf B469} (1996) 387,
{\tt hep-th/9602082}} 
\Rref{F1}{A. Fayyazuddin, {\tt hep-th/9504120}}
\Rref{AFS}{P.C. Argyres, A.E. Faraggi and A.D. Shapere, {\tt hep-th/9505190}}
\Rref{DS2}{U.H. Danielsson and B. Sundborg, Phys. Lett. {\bf B370}
(1996) 83, {\tt hep-th/9511180}}
\Rref{AS}{P.C. Argyres and A.D. Shapere, Nucl. Phys. {\bf B461} (1996)
437, {\tt hep-th/9509175}} 
\Rref{AAG}{M.R. Abolhasani, M. Alishahiha and A.M. Ghezelbash, {\tt
hep-th/9606043}} 
\Rref{DFH}{N. Dorey, C. Fraser, T.J. Hollowood and M.A.C. Kneipp, 
Phys. Lett. {\bf B383} (1996) 422, {\tt hep-th/9605069}}
\Rref{GNO}{P. Goddard, J. Nuyts and D. Olive, Nucl. Phys. {\bf B125}
(1977) 1}
\Rref{SEN}{A. Sen, Phys. Lett. {\bf B329} (1994) 217}
\Rref{POR}{M. Porrati, Phys. Lett. {\bf B377} (1996) 67, {\tt hep-th/9505187}}
\Rref{GL}{J.P. Gauntlett and D. A. Lowe, Nucl. Phys. {\bf B472} (1996) 194,
{\tt hep-th/9601085}}
\Rref{LWY}{K. Lee, E.J. Weinberg and P. Yi, Phys. Lett. {\bf B386}
(1996) 97, {\tt hep-th/9601097}; Phys. Rev. {\bf D54} (1996) 1633, 
{\tt hep-th/9602167}}
\Rref{AGV}{V. Arnold, A. Gusein-Zade and A. Varchenko, {\sl
Singularities of Differentiable Maps. I. II.\/} Birkh\"auser 1985}
\Rref{MUR}{M. Murray, {\tt hep-th/9605054}}
\Rref{GIB1}{G. Gibbons, Phys. 
Lett. {\bf B382} (1996) 53, {\tt hep-th/9603176}}
\Rref{GIB2}{G.W. Gibbons and P. Rychenkova, {\tt hep-th/9608085}}
\Rref{SC}{S.A. Connell, `The Dynamics of the ${\rm SU}(3)$ $(1,1)$
Magnetic Monopole', Unpublished preprint available by anonymous
ftp from : {\tt <ftp://maths.adelaide.edu.au/pure/-
mmurray/oneone.tex>}}
\endref

\ciao